\documentclass[acmtog]{acmart}

\setcopyright{none}
\renewcommand\footnotetextcopyrightpermission[1]{}

\copyrightyear{2026}
\acmYear{2026}

\AtBeginDocument{%
  }

\usepackage{amsmath}
\usepackage{amssymb}
\usepackage{booktabs}
\usepackage{subcaption}
\usepackage{array}
\usepackage{colortbl}
\usepackage{tabularx}
\usepackage{enumitem}
\usepackage{float}

\usepackage{listings}
\usepackage[framemethod=TikZ]{mdframed}

\makeatletter
\def\input@path{{figures/tikz/}}
\makeatother
\usepackage{tikz}
\usetikzlibrary{arrows.meta, calc, decorations.markings}

\definecolor{pathorange}{rgb}{0.902,0.494,0.133}
\definecolor{vertexorange}{rgb}{0.827,0.329,0.000}
\definecolor{surfgray}{rgb}{0.741,0.765,0.780}
\definecolor{adjred}{rgb}{0.890,0.102,0.133}
\definecolor{connblue}{rgb}{0.161,0.502,0.725}
\definecolor{sunouter}{rgb}{0.729,0.557,0.137}
\definecolor{suninner}{rgb}{0.945,0.769,0.059}
\definecolor{jacpurple}{rgb}{0.557,0.267,0.678}
\definecolor{raydark}{rgb}{0.173,0.243,0.314}
\definecolor{framegreen}{rgb}{0.153,0.682,0.376}
\definecolor{framered}{rgb}{0.906,0.298,0.235}
\definecolor{prbblue}{rgb}{0.275,0.510,0.706}
\definecolor{prbvertex}{rgb}{0.122,0.227,0.373}

\newcommand{\lblmain}{\fontsize{8}{9}\selectfont}   % x_i, L_i, \bar L_i
\newcommand{\lblnote}{\fontsize{7}{8}\selectfont}   % per-vertex annotations
\newcommand{\lblhalf}{\fontsize{7}{8}\selectfont}
\newcommand{\lblhalfnote}{\fontsize{6}{7}\selectfont}

\tikzset{
  surface/.style   = {fill=surfgray, draw=black, line width=0.4pt, line join=miter},
  lightpath/.style = {draw=pathorange, line width=1.1pt,
                      -{Stealth[length=5.5pt, width=4.4pt]}, shorten >=1.2pt},
  conn/.style      = {draw=connblue, line width=0.9pt, dash pattern=on 3pt off 2pt,
                      -{Stealth[length=5pt, width=4pt]}},
  adjoint/.style   = {draw=adjred, line width=0.9pt, dash pattern=on 3pt off 2pt,
                      -{Stealth[length=5pt, width=4pt]}},
  faded/.style     = {opacity=0.35},
  lbl/.style       = {font=\lblmain, inner sep=1pt},
  note/.style      = {font=\lblnote, inner sep=1pt},
  hlbl/.style      = {font=\lblhalf, inner sep=1pt},
  hnote/.style     = {font=\lblhalfnote, inner sep=1pt},
  ray/.style       = {draw=raydark, line width=0.8pt, -{Stealth[length=4.5pt, width=3.6pt]}},
  jac/.style       = {draw=jacpurple, line width=0.8pt, dash pattern=on 2.5pt off 2pt,
                      -{Stealth[length=4.5pt, width=3.6pt]}},
  frameu/.style    = {draw=framegreen, line width=0.6pt, -{Stealth[length=2.6pt, width=2.2pt]}},
  framev/.style    = {draw=framered, line width=0.6pt, -{Stealth[length=2.6pt, width=2.2pt]}},
}

\newcommand{\surf}[3]{\path[surface] #1 -- #2 -- #3 -- cycle;}

\newcommand{\vtx}[1]{\fill[vertexorange] #1 circle (1.6pt);}

\newcommand{\lrbsun}[2][1]{%
  \begin{scope}[shift={#2}, scale=#1]
    \fill[suninner!3]  (0,0) circle (58);
    \fill[suninner!6]  (0,0) circle (56);
    \fill[suninner!9]  (0,0) circle (54);
    \fill[suninner!12] (0,0) circle (52);
    \fill[suninner!14] (0,0) circle (50);
    \fill[suninner!17] (0,0) circle (48);
    \path[fill=sunouter, draw=black, line width=0.3pt, rounded corners=0.6pt]
      (0:47.3) -- (9:37.6) -- (18:47.3) -- (27:37.6) -- (36:47.3) -- (45:37.6) -- (54:47.3) -- (63:37.6) --
      (72:47.3) -- (81:37.6) -- (90:47.3) -- (99:37.6) -- (108:47.3) -- (117:37.6) -- (126:47.3) -- (135:37.6) --
      (144:47.3) -- (153:37.6) -- (162:47.3) -- (171:37.6) -- (180:47.3) -- (189:37.6) -- (198:47.3) -- (207:37.6) --
      (216:47.3) -- (225:37.6) -- (234:47.3) -- (243:37.6) -- (252:47.3) -- (261:37.6) -- (270:47.3) -- (279:37.6) --
      (288:47.3) -- (297:37.6) -- (306:47.3) -- (315:37.6) -- (324:47.3) -- (333:37.6) -- (342:47.3) -- (351:37.6) -- cycle;
    \fill[suninner] (0,0) circle (30.6);
  \end{scope}%
}

\newcommand{\screen}[5]{%
  \def\scrxs{#1}%
  \foreach \x/\c in {#1} {
    \ifnum\c=0 \xdef\scrxa{\x}\fi
    \ifnum\c=5 \xdef\scrxb{\x}\fi
  }
  \foreach \x/\c in {#1} {
    \foreach \k in {0,...,5} {
      \pgfmathsetmacro{\yl}{#2+#3*\k}
      \pgfmathsetmacro{\yr}{#4+#5*\k}
      \pgfmathsetmacro{\y}{\yl+(\yr-\yl)/(\scrxb-\scrxa)*(\x-\scrxa)}
      \coordinate (s-\c-\k) at (\x,\y);
    }
  }
  \foreach \c in {0,...,4} {
    \foreach \k in {0,...,4} {
      \pgfmathtruncatemacro{\cc}{\c+1}
      \pgfmathtruncatemacro{\kk}{\k+1}
      \coordinate (px-\c-\k) at ($ (s-\c-\k)!0.5!(s-\cc-\kk) $);
    }
  }
  \path[fill=white] (s-0-0) -- (s-5-0) -- (s-5-5) -- (s-0-5) -- cycle;
}
\newcommand{\hatchpx}[3][]{%
  \pgfmathtruncatemacro{\cc}{#2+1}
  \pgfmathtruncatemacro{\kk}{#3+1}
  \begin{scope}[#1]
    \clip (s-#2-#3) -- (s-\cc-#3) -- (s-\cc-\kk) -- (s-#2-\kk) -- cycle;
    \foreach \d in {-26,-24.2,...,14}
      \draw[connblue, line width=0.5pt] ($ (px-#2-#3) + (\d pt,-12pt) $) -- ++(24pt,24pt);
  \end{scope}%
}
\newcommand{\fillpx}[2]{%
  \pgfmathtruncatemacro{\cc}{#1+1}
  \pgfmathtruncatemacro{\kk}{#2+1}
  \fill[connblue] (s-#1-#2) -- (s-\cc-#2) -- (s-\cc-\kk) -- (s-#1-\kk) -- cycle;%
}
\newcommand{\screenlines}{%
  \foreach \k in {1,...,4} \draw[line width=0.4pt] (s-0-\k) -- (s-5-\k);
  \foreach \c in {1,...,4} \draw[line width=0.4pt] (s-\c-0) -- (s-\c-5);
  \draw[line width=0.7pt, line join=miter] (s-0-0) -- (s-5-0) -- (s-5-5) -- (s-0-5) -- cycle;
}

\definecolor{codebg}{RGB}{248,249,251}
\definecolor{codeaccent}{RGB}{180,190,210}
\definecolor{codecomment}{RGB}{120,130,145}
\definecolor{codekeyword}{RGB}{175,50,115}
\definecolor{codebuiltin}{RGB}{50,115,180}
\definecolor{codestring}{RGB}{140,70,30}
\definecolor{codenumber}{RGB}{150,160,170}

\lstdefinestyle{nicestyle}{
  basicstyle=\ttfamily\small\color{black!85},
  commentstyle=\color{codecomment}\itshape,
  keywordstyle=[1]\color{codekeyword}\bfseries,
  keywordstyle=[2]\color{codebuiltin},
  stringstyle=\color{codestring},
  numberstyle=\tiny\color{codenumber},
  numbers=left,
  numbersep=10pt,
  xleftmargin=20pt,
  frame=none,
  captionpos=b,
  breaklines=true,
  showstringspaces=false,
  tabsize=2,
  mathescape=true,
  language=Python,
  morekeywords=[1]{def,for,if,else,elif,return,break,continue,in,and,or,not,True,False,None,lambda,yield,with,as},
  morekeywords=[2]{range,len,empty_reservoir,intersect,sample_bsdf,connect_to_sensor,forward_jacobian,backward_grad,splat,uniform,consume_reservoir_rng,inv, backward_from, grad, enable_grad, detach_copy, filter_read, read_adjoint, lrb_3pass_backward, reslrb_primal, reslrb_adjoint, indentity, advance_path, identity},
}

\mdfdefinestyle{nicebox}{
  backgroundcolor=codebg,
  linecolor=codeaccent,
  linewidth=0.6pt,
  topline=true,
  bottomline=true,
  rightline=true,
  leftline=true,
  innerleftmargin=1pt,
  innerrightmargin=1pt,
  innertopmargin=1pt,
  innerbottommargin=1pt,
  skipabove=0.5em,
  skipbelow=0.5em,
  roundcorner=6pt,
}

\newif\ifarxiv
\arxivtrue

\begin{document}

\title{Constant-Memory Differentiable Light Tracing}

\author{Linas Beresna}
\orcid{0009-0004-7102-324X}
\affiliation{%
  \institution{Simon Fraser University}
  \department{School of Computing Science}
  \city{Burnaby}
  \state{BC}
  \country{Canada}}
\email{linas\_beresna@sfu.ca}
 
\author{Eugene Fiume}
\orcid{0000-0001-5939-7055}
\affiliation{%
  \institution{Simon Fraser University}
  \department{School of Computing Science}
  \city{Burnaby}
  \state{BC}
  \country{Canada}}
\email{eugene\_fiume@sfu.ca}

\renewcommand{\shortauthors}{Beresna and Fiume}

\begin{abstract}
Reverse-mode differentiation of Monte Carlo light transport naively stores a computation graph that grows with path length, ruling it out for deep or high-sample simulation. Path Replay Backpropagation (PRB) removes that cost for viewpoint path tracing by reconstructing paths from their seeds and replaying them backward in constant memory. The construction rests on a one-path-one-pixel property that light tracing lacks: one light path contributes to many pixels, and its per-vertex adjoints cannot be recovered during replay from constant-size state.

We show this prevents a direct extension of PRB, and that buffering the per-vertex data stays linear in path length. We then evaluate two constant-memory formulations. Reservoir Light Replay Backpropagation compresses a path's sensor connections into one representative by weighted reservoir sampling, keeping two passes at the cost of a sampling decision. LRB-3-pass adds a traversal that accumulates downstream adjoints first, keeping every connection and matching naive automatic differentiation in expectation.

Both support differentiating with fixed or parameter-dependent path sampling (detached and attached), and both extend to participating media. We validate against naive AD and finite differences, characterize the memory and runtime trade-offs, and recover specular and volumetric caustics by light tracing alone.
\end{abstract}



\begin{teaserfigure}
  \includegraphics[width=\textwidth]{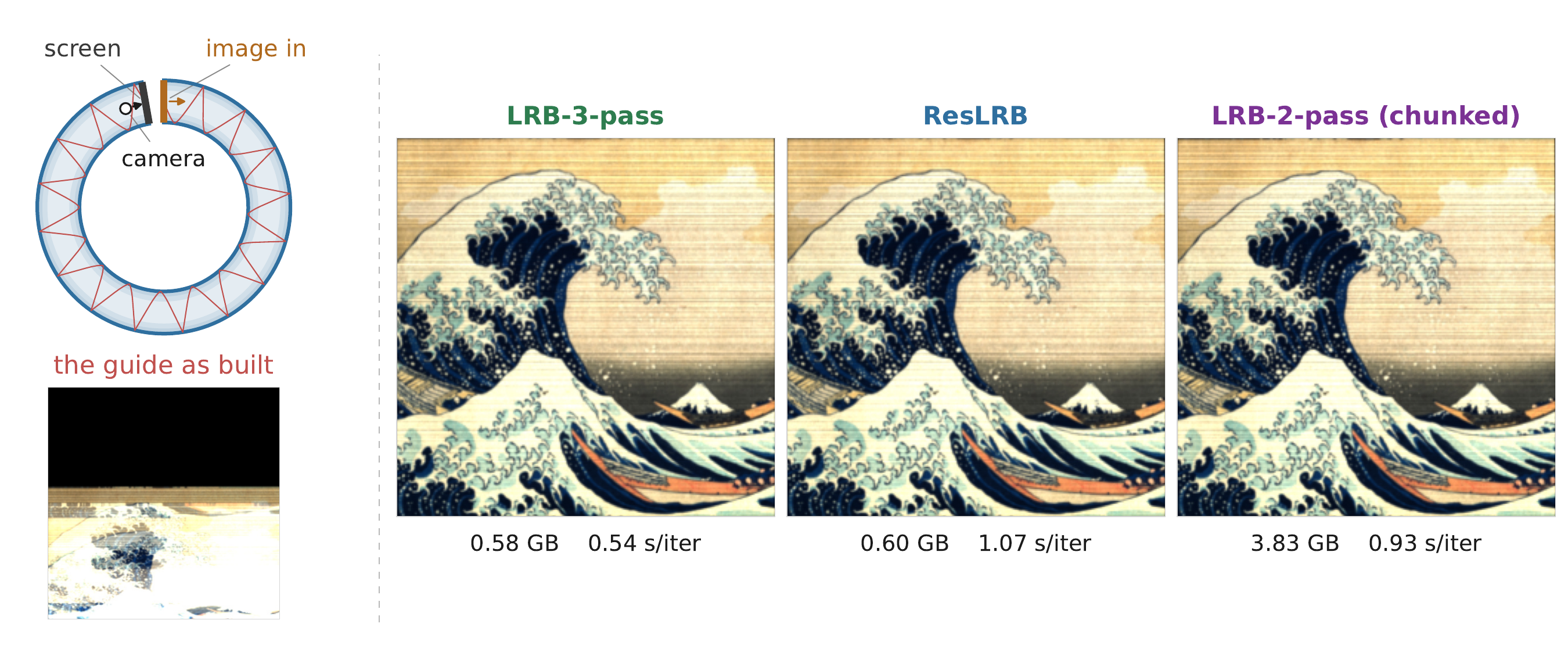}
  \caption{A glass light guide coiled through 350 degrees, where no light crosses the bend in fewer than 14 reflections.
  A wall tilt of a thousandth of a degree steers the light off the screen and takes the image with it
  (\emph{left}). We recover the wall profile from the ruined image through 32 and 36 reflections with
  LRB-3-pass, ResLRB and chunked LRB-2-pass (\emph{right}), shown with peak memory and time per iteration.
  The two replay formulations store nothing per vertex; LRB-2-pass buffers every vertex and needs over six times the memory.}
  \label{fig:teaser}
\end{teaserfigure}

\maketitle

\ifarxiv
\pagestyle{plain}
\thispagestyle{plain}
\fi

\section{Introduction}
\label{sec:introduction}

Differentiable rendering has become a core tool in computer graphics and vision, recovering scene parameters from images through gradient-based optimization.
The parameter spaces involved are large, and could include objects such as: displacement maps, spatially varying BRDFs, or meshes with thousands of vertices. As such, reverse-mode automatic differentiation (AD) is the practical choice: one backward pass produces gradients for every parameter at once.
The cost is memory. The computation graph records every scattering event along every path, so it grows with both sample count and path length, and deep transport at high sample counts can easily exhaust GPU memory.

Path Replay Backpropagation (PRB)~\cite{Vicini2021PathReplay} removes that cost for viewpoint path tracing by reconstructing paths from their pseudorandom seeds and differentiating each vertex locally during a backward replay.
Because each path contributes to a single pixel, the image-space adjoint it needs is one scalar, known before the replay begins, and memory is constant in path length.

\begin{figure}[t]
  \centering
  \begin{subfigure}[b]{0.48\columnwidth}
    \centering
    \includegraphics[width=\linewidth]{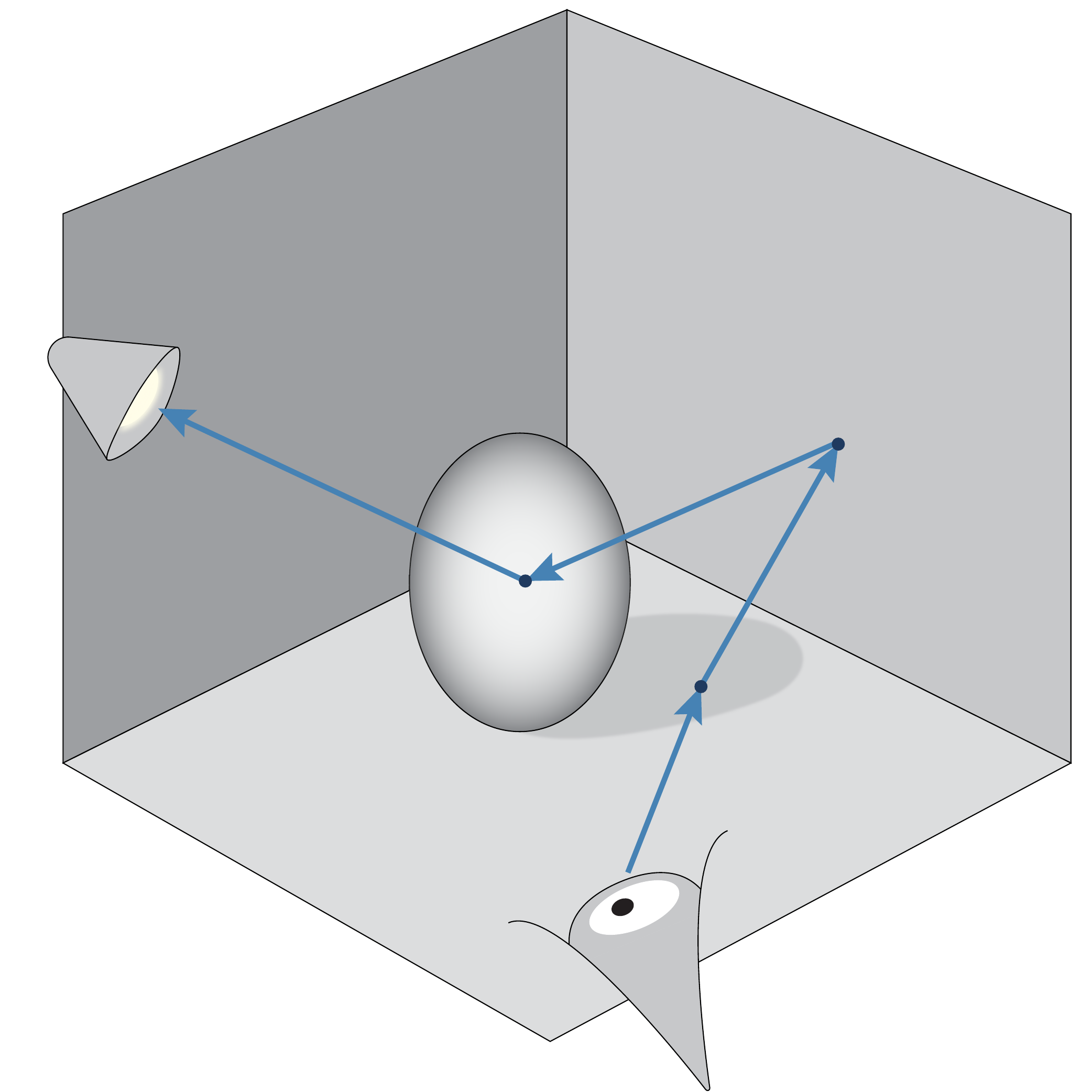}
    \caption{Viewpoint Path Tracing}
    \label{fig:path-tracing}
  \end{subfigure}
  \hfill
  \begin{subfigure}[b]{0.48\columnwidth}
    \centering
    \includegraphics[width=\linewidth]{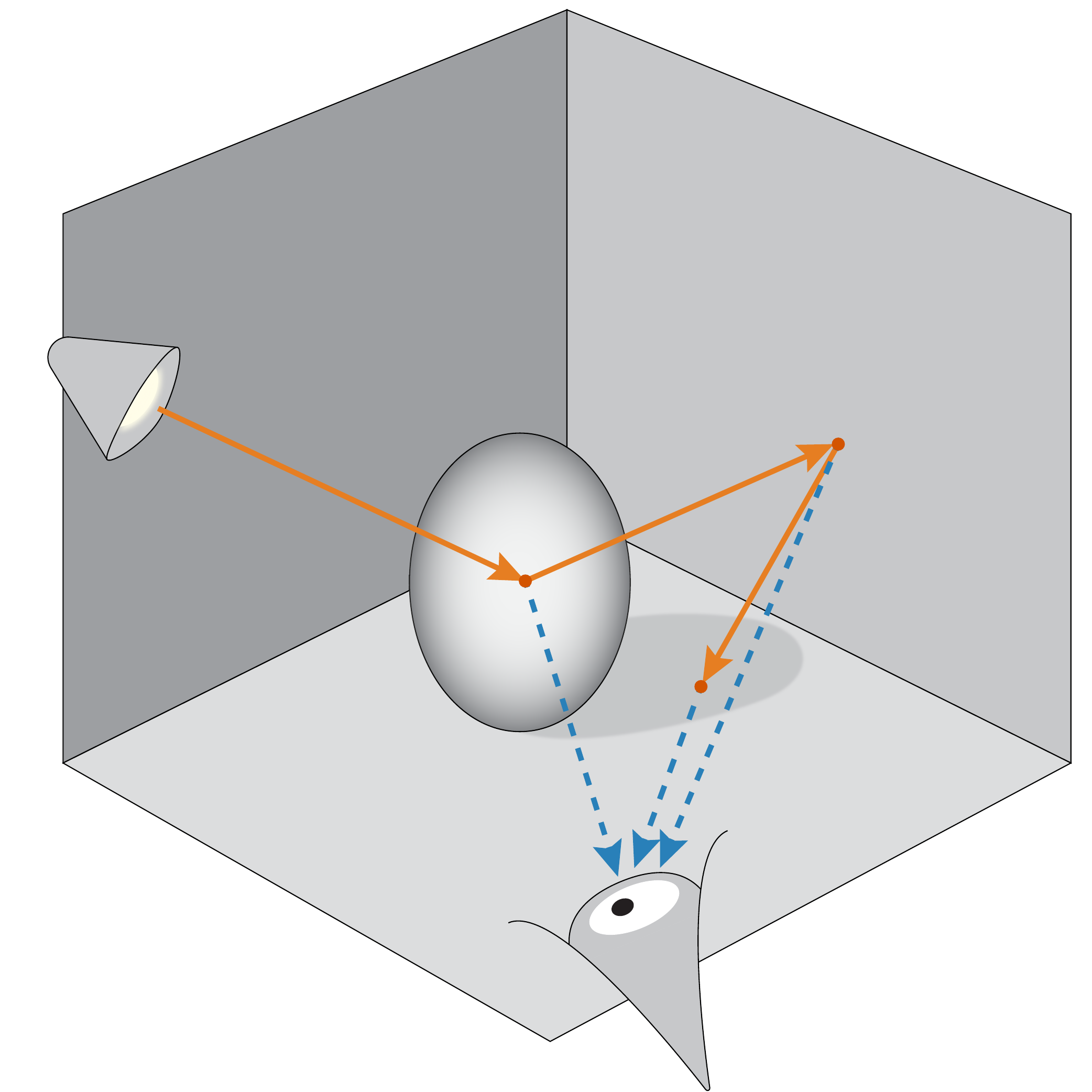}
    \caption{Light Tracing}
    \label{fig:light-tracing}
  \end{subfigure}
  \caption{Comparison of path--sensor connectivity. In viewpoint path tracing (a), a path from the sensor typically contributes to the image via a single emitter connection. In light tracing (b), a path from the emitter can connect to the sensor at every scattering vertex (dashed blue arrows), causing the computation graph to grow with both path length and the number of valid connections.}
  \label{fig:connectivity-comparison}
\end{figure}

We would like the same property for light tracing, which propagates paths outward from emitters.
At each scattering event the renderer can connect to the sensor and splat flux, so a path of depth $k$ may make $k$ contributions involving up to $k$ pixels (Figure~\ref{fig:connectivity-comparison}).
That many-to-many connectivity is what makes light tracing effective for concentrated illumination such as caustics and specular chains, which camera-side tracers must work hard to discover.

The adjoint pass must weight each replayed vertex by how the loss responds to the flux splatted there.
With one pixel per path that weight is a single scalar, fixed in advance; with many, the per-vertex adjoints depend on sensor connections further along the path and cannot be recovered during a forward replay from constant-size state.
The direct remedy is to keep PRB's two passes and buffer what the forward replay cannot recover.
This \emph{LRB-2-pass} variant records every splat position and radiance in the primal pass and reads them back during the adjoint pass; its gradients match naive AD, but the buffer grows with path length, so memory stays linear.

Two methods remove the obstacle, each paying for constant memory differently.
\emph{Reservoir Light Replay Backpropagation} (ResLRB)~\cite{Beresna2026Stochastic} keeps PRB's two passes by compressing the per-path sensor connections into a single stochastically selected representative.
Each valid connection is streamed into a weighted reservoir during the primal pass; the adjoint pass replays to the selected depth and backpropagates through the survivor, reweighted by the inverse selection probability so the estimate stays unbiased.
Here, we introduce \emph{LRB-3-pass}, which keeps every connection by adding a traversal.
A first pass splats as usual and the image-space loss gradient is stored as a lookup tensor; a second re-traces and accumulates the adjoint-weighted radiance along each path; a third consumes that sum by subtraction while backpropagating at each vertex.
Only fixed-size accumulators are carried, so memory is constant and the gradient matches naive AD in expectation.

Both methods support detached and attached formulations.
Specifically, the detached case covers parameters that change scattering values without moving the path, such as material albedo or emitter intensity.
The attached case covers parameters that move the vertices themselves, which is required for geometry optimization in specular transport, such as a lens surface whose shape sets a caustic.
\citet{Beresna2026Stochastic} sketch the attached extension of ResLRB; we give the attached formulation in full for both methods.

Our contributions include:
\begin{itemize}[leftmargin=*, topsep=1pt]
\item \emph{Obstruction evaluation.} We show why PRB's two-pass replay cannot achieve constant memory for light tracing, and that the buffered LRB-2-pass stays linear in path length.
\item \emph{LRB-3-pass.} A third traversal makes differentiable light tracing constant-memory without discarding a single sensor connection, matching naive AD in expectation, with detached and attached formulations.
\item \emph{Trade-off.} Naive AD, LRB-2-pass, ResLRB, and LRB-3-pass are evaluated side by side on memory, variance, and runtime, and both constant-memory methods recover specular and volumetric caustics by light tracing alone.
\end{itemize}

\section{Related Work}
\label{sec:related}

This section situates our work against prior methods for adjoint differentiable rendering, differentiable light transport beyond path tracing, and reservoir-based sampling in rendering.

\subsection{Memory-Efficient Adjoint Methods}
Naive reverse-mode AD records every operation of the forward simulation, requiring memory that grows with path length and sample count.
Radiative Backpropagation~\cite{NimierDavid2020Radiative} reformulates gradient computation as a separate adjoint transport problem, reducing memory at the cost of quadratic sample complexity: estimating the adjoint radiance at each primal path vertex requires an independent Monte Carlo simulation, making the overall cost proportional to the product of primal and adjoint sample counts.
PRB~\cite{Vicini2021PathReplay} eliminates both the memory growth and the quadratic cost for viewpoint path tracing by replaying paths from pseudorandom number generator (PRNG) seeds and backpropagating locally at each vertex.
The replay framework has since been extended to non-static geometry~\cite{worchel2025radiative}, time-resolved acoustic transport~\cite{FinnendahlAcoustic2025}, and branching random walks in Monte Carlo PDE solvers~\cite{Yilmazer2024Solving}.

\subsection{Differentiable Rendering beyond Path Tracing}
Camera-side path tracing and its variants have been the primary focus of physics-based differentiable rendering~\cite{LiEdge2018, Loubet2019Reparameterizing, BangaruWarp2020, Vicini2022sdf, Zhang2020Path,Zhang2021Media, Yan2024Variance, Zhang2026Cache}, but several recent methods have broadened the scope.
Mitsuba~2's particle tracer (\texttt{ptracer})~\cite{NimierDavidVicini2019Mitsuba2} provides a differentiable light tracer via naive reverse-mode AD, with memory that grows linearly in path length.
\citet{Lipp2024Light} describe an adjoint light tracer for lighting design that targets object-space irradiance rather than image-space derivatives.
\citet{nimierdavid2022unbiased} use reservoir sampling within differentiable volume rendering to select a single distance along a ray for transmittance gradient evaluation, yielding an unbiased single-sample estimator structurally similar to ResLRB but applied within a volume rather than along a surface light path.
\citet{Zeltner2021MonteCarlo} develop a taxonomy of Monte Carlo estimators for differentiable light transport, analyzing how sampling strategy choices affect gradient variance in both detached and attached settings.
\citet{fischer2023plateau} convolve the rendering function with a kernel to capture long-range parameter-to-pixel relationships at the cost of increased variance in high-dimensional parameter spaces.

Extended Path Space Manifolds (EPSMs)~\cite{Xing2023Manifolds} and their extension to differentiable photon mapping (DPM)~\cite{Xing2024DPMG} are closest in motivation to our work EPSMs compute geometric derivatives through specular chains, and DPM-G adds contribution derivatives, enabling optimization of specular-diffuse-specular (SDS) transport via photon mapping. DPM addresses a sampling problem: SDS paths that unidirectional tracers cannot reach, and bounds backward-pass memory by building gradient paths from a subset of pixels per iteration; we address a memory problem, and process every pixel and sample, with the bound coming from the structure of the replay.

\subsection{Reservoir Sampling and Resampling}
Weighted reservoir sampling~\cite{Vitter1985Res} selects samples from a stream of unknown length in a single pass.
Resampled Importance Sampling~\cite{Talbot2005Global} introduced streaming resampling for variance reduction in rendering, and ReSTIR~\cite{1Bitterli2020Restir} applied spatiotemporal reservoir reuse to direct lighting.
Generalized Resampled Importance Sampling~\cite{Lin2022ReSTIR} extended the framework to select a single representative path from a larger path tree, and ReSTIR BDPT~\cite{Hedstrom2025BDPT} applies similar ideas to bidirectional path tracing including the selection of light-to-sensor connections.
In differentiable rendering, Parameter-space ReSTIR~\cite{Chang2023ReSTIR} and amortized inverse rendering~\cite{Wang2023ReSTIR} use reservoir-based techniques for sample reuse during optimization.
The ResLRB method~\cite{Beresna2026Stochastic} uses a weighted reservoir to compress the set of per-path sensor connections to a single representative, serving as a mechanism for bounding the AD graph.
The LRB-3-pass method does not use reservoir sampling; it retains all connections through multi-pass accumulation.

\section{Background}
\label{sec:background}

This section establishes the notation used throughout the paper and summarizes the two technical ingredients on which both methods build: the structure of light tracing with per-vertex sensor connections, and the path replay formulation of \citet{Vicini2021PathReplay} for constant-memory reverse-mode differentiation.

\subsection{Light Tracing}
\label{sec:background-light-tracing}

A light tracer generates paths starting from emitters and propagates them into the scene.
A path $\bar{x} = (x_0, x_1, \dots, x_k)$ begins at a point $x_0$ sampled on an emitter surface and scatters through the scene via importance-sampled BSDF interactions at each subsequent vertex $x_i$.
At every scattering vertex $x_i$ for $i \geq 1$, the renderer tests whether $x_i$ can be connected to the sensor: if the connection is unoccluded and the surface is non-specular (fully specular BSDFs have zero contribution in any fixed direction and cannot be evaluated for an arbitrary sensor connection), the path contributes flux to a pixel.

Following the path integral formulation of \citet{veach1998robust}, the measurement contribution of the full path to the image is
\begin{equation}
I(\bar{x}) = \sum_{i=1}^{k} W_e(x_i \to x_s)\, f_i(\bar{x})\, G(x_i \leftrightarrow x_s)\, V(x_i \leftrightarrow x_s),
\label{eq:light-path-measurement}
\end{equation}
where $x_s$ denotes the point on the sensor reached by the connection from $x_i$, $W_e$ is the sensor importance (emitted importance response), $G$ is the geometric term between $x_i$ and $x_s$, $V$ is a binary visibility term, and $f_i$ collects the path throughput from the emitter through the first $i$ scattering events:
\begin{equation}
f_i(\bar{x}) = L_e(x_0 \to x_1) \prod_{j=1}^{i} f_s(x_{j-1} \to x_j \to x_{j+1}^{*})\, G(x_{j-1} \leftrightarrow x_j),
\label{eq:path-throughput}
\end{equation}
where $L_e$ is the emitted radiance at the source, $f_s$ is the BSDF, and $x_{j+1}^{*}$ is either $x_{j+1}$ (the next path vertex) when $j < i$, or $x_s$ (the sensor point) when $j = i$.
The product telescopes: each additional vertex appends one BSDF evaluation and one geometric term to the running throughput.

Each non-zero term in the sum of Eq. \eqref{eq:light-path-measurement} corresponds to a sensor connection that lands at some coordinate $u_i$ on the image plane.
The rendered image is formed by splatting each contribution $L_i = W_e \, f_i \, G \, V$ into the image through a reconstruction filter centered at $u_i$.
A path of length $k$ can therefore produce up to $k$ splats, each at a potentially different pixel, and the per-path contribution to the image is the multiset $\{(L_i, u_i)\}_{i=1}^{k}$.

This stands in contrast to viewpoint path tracing, where a path originates at the sensor and is connected to an emitter, either by BSDF sampling or next-event estimation (NEE), producing a single contribution to a single pixel.

\subsection{Path Replay Backpropagation}
\label{sec:background-prb}

Given a scalar loss $\mathcal{L}$ defined on the rendered image, the goal of reverse-mode differentiable rendering is to compute $\nabla_\theta \mathcal{L}$, the gradient of the loss with respect to a set of scene parameters $\theta$.
Naive reverse-mode AD records the full computation graph of the forward simulation and traverses it in reverse.
The memory cost scales with the total number of recorded operations, which in a Monte Carlo renderer grows with both path length and sample count.
PRB eliminates the depth dependence for viewpoint path tracing by observing that the path can be reconstructed exactly from the pseudorandom seed that generated it, so there is no need to store the graph, assuming one is willing to pay the price of a second traversal.

\begin{figure}[t]
  \centering
  \begin{subfigure}[b]{0.48\columnwidth}
    \centering
    % PRB adjoint (Figure 3a): one pixel, one scalar adjoint \bar L for the
% whole path. Scaling terms omitted, as stated in the caption.
% Input directly by main.tex; prb-adjoint.tex wraps it to build the PDF.
\begin{tikzpicture}[x=0.1167pt, y=0.1167pt]
  \coordinate (X1) at (739.3,292.7);
  \coordinate (X2) at (239.3,581.2);
  \coordinate (X3) at (858.9,724.5);
  \useasboundingbox (0,0) rectangle (1000,1000);

  \screen{39.6/0,81.5/1,127/2,176.7/3,231.1/4,290.9/5}{111.15}{50.45}{16.5}{63.07}
  \fillpx{3}{2}
  \screenlines

  \surf{(305.3,738.9)}{(68.8,472.3)}{(343.7,532.5)}
  \surf{(652.4,435.6)}{(678.6,141.4)}{(947.9,258.1)}
  \surf{(963.5,841.5)}{(726.7,745)}{(886.6,587.1)}
  \lrbsun[1.52]{(426.4,911.5)}

  \tikzset{prbray/.style={draw=prbblue, line width=0.8pt,
           -{Stealth[length=4.5pt, width=3.6pt]}, shorten >=1pt}}
  \draw[prbray] (203.7,194.7) -- (X1);
  \draw[prbray] (X1) -- (X2);
  \draw[prbray] (X2) -- (X3);
  \draw[prbray] (X3) -- (511.4,861.4);
  \foreach \p in {X1,X2,X3} \fill[prbvertex] (\p) circle (1.3pt);

  \node[hlbl, anchor=base west] at (760,236) {$x_1$};
  \node[hlbl, anchor=base east] at (226,540) {$x_2$};
  \node[hlbl, anchor=base west] at (926,628) {$x_3$};
  \node[hlbl, anchor=base west] at (318,236) {$\bar{L}$};

  \node[hnote, anchor=base west] at (700,452) {$\bar{L} \cdot \nabla_\theta f_s^{(1)}$};
  \node[hnote, anchor=base west] at (12,752)  {$\bar{L} \cdot \nabla_\theta f_s^{(2)}$};
  \node[hnote, anchor=base west] at (640,880) {$\bar{L} \cdot \nabla_\theta f_s^{(3)}$};
\end{tikzpicture}
    \caption{PRB (path tracing)}
    \label{fig:prb-adjoint}
  \end{subfigure}
  \hfill
  \begin{subfigure}[b]{0.48\columnwidth}
    \centering
    % Light-tracing adjoint (Figure 3b): every vertex splats to its own
% pixel, so the weight at x_j needs the adjoints of all connections at or
% after x_j. Vertices are numbered from the emitter, as in the rest of
% the paper. Scaling terms omitted, as stated in the caption.
% Input directly by main.tex; ptracer-adjoint.tex wraps it to build the PDF.
\begin{tikzpicture}[x=0.1167pt, y=0.1167pt]
  \coordinate (X1) at (858.9,724.5);
  \coordinate (X2) at (239.3,581.2);
  \coordinate (X3) at (739.3,292.7);
  \useasboundingbox (0,0) rectangle (1000,1000);

  \screen{39.6/0,81.4/1,126.8/2,176.4/3,230.6/4,290.3/5}{111}{50.32}{16.6}{62.9}
  \hatchpx{4}{4}\hatchpx{1}{3}\hatchpx{3}{1}
  \screenlines

  \surf{(305.3,738.9)}{(68.8,472.3)}{(343.7,532.5)}
  \surf{(652.4,435.6)}{(678.6,141.4)}{(947.9,258.1)}
  \surf{(963.5,841.5)}{(726.7,745)}{(886.6,587.1)}
  \lrbsun[1.52]{(426.4,911.5)}

  \tikzset{lp/.style={lightpath, line width=0.8pt, -{Stealth[length=4.5pt, width=3.6pt]}},
           cn/.style={conn, line width=0.7pt, dash pattern=on 2.2pt off 1.6pt,
                      -{Stealth[length=4pt, width=3.2pt]}}}
  % sensor connections
  \draw[cn, shorten <=2pt] (X1) -- (px-4-4);
  \draw[cn, shorten <=2pt] (X2) -- (px-1-3);
  \draw[cn, shorten <=2pt] (X3) -- (px-3-1);
  % the light path, from the emitter
  \draw[lp] (511.4,861.4) -- (X1);
  \draw[lp] (X1) -- (X2);
  \draw[lp] (X2) -- (X3);
  \foreach \p in {X1,X2,X3} {\vtx{(\p)}}

  % adjoints flow back along each connection
  \tikzset{adj/.style={draw=adjred, line width=0.7pt, -{Stealth[length=3.5pt, width=2.8pt]}}}
  \draw[adj] (300.2,352.2) -- (418.7,438.7);
  \draw[adj] (125,362.7) -- (167.9,449.2);
  \draw[adj] (311.3,184.9) -- (449.6,225.7);
  \node[hlbl, anchor=base east] at (290,400) {$\bar{L}_1$};
  \node[hlbl, anchor=base east] at (112,392) {$\bar{L}_2$};
  \node[hlbl, anchor=base west] at (330,232) {$\bar{L}_3$};

  \node[hlbl, anchor=base west] at (926,628) {$x_1$};
  \node[hlbl, anchor=base east] at (226,540) {$x_2$};
  \node[hlbl, anchor=base west] at (760,236) {$x_3$};

  \node[hnote, anchor=base east] at (995,890)
    {$\big(\sum_{i=1}^{3} \bar{L}_i\big) \cdot \nabla_\theta f_s^{(1)}$};
  \node[hnote, anchor=base west] at (8,770)
    {$\big(\sum_{i=2}^{3} \bar{L}_i\big) \cdot \nabla_\theta f_s^{(2)}$};
  \node[hnote, anchor=base west] at (700,452)
    {$\bar{L}_3 \cdot \nabla_\theta f_s^{(3)}$};
\end{tikzpicture}
    \caption{Light Tracing}
    \label{fig:lt-adjoint}
  \end{subfigure}
  \caption{The adjoint structure of PRB and its obstruction in light tracing. In path tracing (a), the path contributes to one pixel, so a single scalar adjoint $\bar{L}$ can be precomputed before replay. In light tracing (b), each vertex $x_i$ splats to a different pixel, requiring per-vertex adjoints $\bar{L}_i$ that depend on future splat locations not yet visited during a forward replay. Scaling terms are omitted for brevity.}
  \label{fig:adjoint-obstruction}
\end{figure}

The algorithm proceeds in two passes.
In the primal pass, a path is traced from the sensor into the scene, its radiance contribution $L$ is evaluated and accumulated into the image, and two quantities are recorded per sample: the PRNG seed used to generate the path, and the final radiance value $L$, whilst no AD graph is retained.

In the adjoint pass, shown in Figure~\ref{fig:prb-adjoint}, the PRNG is re-seeded and the path is replayed from scratch.
At each vertex $x_i$ of the replayed path, reverse-mode AD is enabled over a local scope that covers only the scattering computation at that vertex.
The gradient contribution from vertex $x_i$ is computed using the log-derivative identity: rather than differentiating the full product of BSDF evaluations from $x_0$ to $x_k$, the cached radiance $L$ is used as a multiplicative weight and only the local BSDF factor $f_s^{(i)}$ is differentiated.
Concretely, the adjoint at each vertex takes the form
\begin{equation}
\bar{L} \cdot L \cdot \nabla_\theta \log f_s^{(i)},
\label{eq:prb-log-deriv}
\end{equation}
where $\bar{L} = \partial \mathcal{L} / \partial L$ is the image-space adjoint at the pixel this path contributes to.
The local factor enters only through $\nabla_\theta \log f_s^{(i)} = \nabla_\theta (f_s^{(i)}) / f_s^{(i)}$, and the cached radiance $L$ carries the contribution of all other vertices, so each vertex is differentiated independently.
After the local gradient is propagated into the scene parameters, the AD graph for that vertex is discarded before the next vertex is processed.

\subsection{Detached and Attached Derivatives}
\label{sec:background-attached}

The gradient $\nabla_\theta \mathcal{L}$ depends on how the scene parameters $\theta$ influence the path contribution.
There are two fundamentally different channels through which this influence can act.

In the \emph{detached} case, the parameters affect the values computed along the path, such as BSDF reflectance, emitter radiance, or sensor importance, but not the positions of the path vertices themselves.
The path geometry is treated as fixed with respect to $\theta$: the vertices $x_0, \dots, x_k$ and the sensor coordinates $u_i$ do not move when $\theta$ is perturbed.
Only the scattering weights and the emitted radiance respond to the perturbation.
This is the setting described in the previous subsection, and is handled directly by the log-derivative identity of Eq. \eqref{eq:prb-log-deriv}.

In the \emph{attached} case, the parameters also affect the path geometry.
A perturbation of a surface, for instance, shifts the intersection point $x_i$ on that surface, which changes the incident and outgoing directions at $x_i$, which in turn shifts the next intersection point $x_{i+1}$, and so on down the path.
A single parameter perturbation can therefore propagate through every subsequent link in the path, altering both the scattering values and the positions at which they are evaluated.
In the terminology of \citet{Vicini2021PathReplay}, the samples are \emph{attached} to the underlying geometry: they move with it rather than remaining fixed.

\begin{figure}[t]
	\centering
	% Attached ray parameterization (Figure 4): segments p_2 = (v_1, v_2) and
% p_3 = (v_2, v_3), and the local ray Jacobian J^ray_2 = dp_3/dp_2
% (Eq. local-ray-jacobian).
% Input directly by main.tex; attached-param.tex wraps it to build the PDF.
\begin{tikzpicture}[x=0.2571pt, y=0.2571pt]
  \coordinate (V1) at (135.5,126.8);
  \coordinate (V2) at (500,408);
  \coordinate (V3) at (822,155);
  \useasboundingbox (15,0) rectangle (960,505);

  \surf{(25.7,245.3)}{(64.6,8.3)}{(309.7,105)}
  \surf{(500.2,493.8)}{(291,362.1)}{(665.1,328.8)}
  \surf{(857,310)}{(689.4,59.8)}{(946.1,96.6)}

  % local surface frames
  \draw[frameu] (V1) -- ++(0,40);  \draw[framev] (V1) -- ++(-40,0);
  \draw[frameu] (V2) -- ++(0,40);  \draw[framev] (V2) -- ++(-40,0);
  \draw[frameu] (V3) -- ++(0,40);  \draw[framev] (V3) -- ++(40,0);

  \draw[ray, shorten >=2.6pt] (V1) -- (V2);
  \draw[ray, shorten >=2.6pt] (V2) -- (V3);
  \draw[jac] (210.4,116.8) .. controls (260.5,228.7) and (368.1,302.4) .. (487.6,306.5)
             .. controls (601.4,310.5) and (708.7,250.9) .. (772.7,143.7);
  \foreach \p in {V1,V2,V3} {\fill[vertexorange] (\p) circle (2.4pt);}

  \node[lbl, anchor=base west] at (146,88)  {$\mathbf{v}_1$};
  \node[lbl, anchor=base west] at (511,404) {$\mathbf{v}_2$};
  \node[lbl, anchor=base west] at (823,112) {$\mathbf{v}_3$};
  \node[lbl, anchor=base west] at (244,272) {$\mathbf{p}_2$};
  \node[lbl, anchor=base west] at (700,272) {$\mathbf{p}_3$};
  \node[lbl] at (487,248)
    {$J^{\text{ray}}_2 = \dfrac{\partial \mathbf{p}_3}{\partial \mathbf{p}_2}$};
\end{tikzpicture}
	\caption{Attached ray parameterization. The local ray Jacobian $J^{\text{ray}}_2$ is the partial derivative of the ray segment $\mathbf{p}_3$ with respect to $\mathbf{p}_2$. Each ray segment $\mathbf{p}_i$ is parameterized by the surface coordinates $\mathbf{v}_{i-1}$ and $\mathbf{v}_{i}$ of its endpoints.}
 	\label{fig:attached-param}
\end{figure}  

The challenge is to track this propagation without recording the full computation graph from $x_0$ to $x_k$.
PRB handles this using forward-mode differentiation over a local parameterization of each ray segment, accumulated multiplicatively along the path.
Specifically, each ray segment ending at vertex $x_i$ is parameterized by the surface coordinates of its two endpoints,
\begin{equation}
\mathbf{p}_i = (\mathbf{v}_{i-1},\, \mathbf{v}_i) \in \mathbb{R}^4,
\label{eq:two-point-param}
\end{equation}
where $\mathbf{v}_j \in \mathbb{R}^2$ are the barycentric (or local UV) coordinates of $x_j$ on the mesh it lies on.
At each vertex during the primal pass, forward-mode AD with respect to $\mathbf{p}_i$ yields a local Jacobian
\begin{equation}
J_i^{\text{ray}} = \frac{\partial \mathbf{p}_{i+1}}{\partial \mathbf{p}_i} \in \mathbb{R}^{4 \times 4},
\label{eq:local-ray-jacobian}
\end{equation}
which describes how a perturbation of the current ray segment shifts the next one. Figure~\ref{fig:attached-param} illustrates an isolated calculation of one ray Jacobian.
The chain rule gives the accumulated Jacobian from the first segment to the $i$-th,
\begin{equation}
J^{\text{ray}}_{0 \to i} = J_{i-1}^{\text{ray}} \cdot J_{i-2}^{\text{ray}} \cdots J_0^{\text{ray}},
\label{eq:accumulated-jacobian}
\end{equation}
which is built incrementally during the primal pass by a single matrix multiplication per vertex, requiring only one $4 \times 4$ matrix of running storage.

During the adjoint pass, the accumulated Jacobian serves as a bridge between frames of reference.
The adjoint arriving from the loss is expressed in terms of the selected vertex's parameterization, but the local scattering computation at an intermediate vertex $x_j$ is expressed in terms of $\mathbf{p}_j$.
Projecting from one to the other requires the inverse of the accumulated Jacobian up to that point:
\begin{equation}
\frac{\partial (\cdot)}{\partial \mathbf{p}_j} = \frac{\partial (\cdot)}{\partial \mathbf{p}_0} \cdot \left(J^{\text{ray}}_{0 \to j}\right)^{-1}.
\label{eq:jacobian-projection}
\end{equation}
The accumulated Jacobian $J^{\text{ray}}_{0 \to j}$ is rebuilt during replay by the same forward-mode process used in the primal pass.

In viewpoint path tracing the path starts at the sensor, so the pixel coordinate is fixed by the camera ray and the accumulated Jacobian carries only the radiance adjoint.
In light tracing the path starts at the emitter, and a perturbation at vertex $x_j$ propagates forward and changes the sensor coordinate $u_i$ at which a downstream vertex $x_i$ splats.
The adjoint must therefore track the sensitivity of the loss to both the splatted radiance $L_i$ and the splatted position $u_i$.
Section~\ref{sec:method} carries this second channel, the position adjoint $\partial \mathcal{L} / \partial u_i$, through the replay.

\section{Method}
\label{sec:method}

This section develops LRB-3-pass and places it alongside the two-pass alternatives.
We begin by examining why PRB's two-pass structure does not generalize to light tracing in constant memory, and why buffering the missing per-vertex data, as LRB-2-pass does, leaves memory linear in path length (Section~\ref{sec:two-pass-obstruction}).
ResLRB (Section~\ref{sec:reslrb}) keeps two passes by retaining a single stochastically selected sensor connection per path; we restate it in the notation used throughout.
LRB-3-pass (Section~\ref{sec:lrb-3pass}) instead retains every connection through a third traversal and matches naive AD in expectation.

\subsection{The Two-Pass Obstacle}
\label{sec:two-pass-obstruction}

Suppose we attempt the most straightforward formulation of PRB to light tracing: trace light paths and splat their contributions in a primal pass, record the PRNG seed and total splatted flux, then replay each path from its seed and backpropagate at each vertex.

The replay reconstructs the path geometry exactly, so we can revisit every vertex $x_i$ and re-evaluate the local scattering computation with AD enabled.
At each vertex we need to weight the local gradient by the image-space adjoint $\bar{L}_i = \partial \mathcal{L} / \partial L_i$, which depends on which pixel $u_i$ the splat landed on.
Unlike PRB, where a single scalar $\bar{L}$ serves the entire path, different vertices along a light path splat to different pixels, so there is no single precomputable adjoint (Figure~\ref{fig:adjoint-obstruction}).

The per-vertex adjoint can be looked up from the image-space gradient tensor after the primal pass, since $u_i$ is reconstructible during replay. However, the difficulty is that a perturbation at vertex $x_j$ affects all subsequent sensor connections along the path. As a result, the gradient at $x_j$ must be weighted by the sum of adjoint-weighted contributions from all downstream vertices. During a forward replay, these downstream contributions have not yet been visited, so the required weighting cannot be computed without storing per-vertex information.

Resolving this within a two-pass framework requires storing per-vertex data, such as splat positions or adjoint-weighted contributions.

Splitting the samples into chunks that are backpropagated and accumulated in turn bounds the peak memory, since each chunk holds only its own share of the per-vertex data. It does not remove the scaling. The storage per sample per vertex is unchanged, so the number of chunks needed grows linearly with path length.

\begin{figure}[t]
  \centering
  \begin{subfigure}[b]{\columnwidth}
    \centering
    % ResLRB, primal pass: the reservoir keeps one sensor connection (here
% at x2), which is splatted reweighted by Omega_k / omega_R
% (Eq. reservoir-estimator). Discarded candidates are faded.
% Input directly by main.tex; reslrb-pass1.tex wraps it to build the PDF.
\begin{tikzpicture}[x=0.2531pt, y=0.2531pt]
  \coordinate (X1) at (725.7,63.9);
  \coordinate (X2) at (518.2,322.2);
  \coordinate (X3) at (357.5,70.6);
  \coordinate (E)  at (873.4,282.3);
  \useasboundingbox (40,0) rectangle (1000,395);

  \screen{48.5/0,75.5/1,104.9/2,136.9/3,171.9/4,210.5/5}{131.2}{32.5}{70.2}{40.64}
  \hatchpx{4}{4}
  \hatchpx[faded]{1}{3}\hatchpx[faded]{3}{1}
  \screenlines

  \surf{(837.4,120.7)}{(622,66.8)}{(738.7,12.9)}
  \surf{(517.5,368.8)}{(413.1,304.7)}{(604,282.3)}
  \begin{scope}[faded]\surf{(442.4,75.8)}{(271.8,102.7)}{(358.3,33.4)}\end{scope}
  \lrbsun[1.173]{(909.3,327.2)}

  % discarded candidates
  \begin{scope}[faded]
    \draw[lightpath, dash pattern=on 3pt off 2pt] (X2) -- (X3);
    \draw[lightpath, line width=0.9pt, dash pattern=on 3pt off 2pt] (X1) -- (px-1-3);
    \draw[lightpath, line width=0.9pt, dash pattern=on 3pt off 2pt] (X3) -- (px-3-1);
    \vtx{(X3)}
    \node[lbl, anchor=base west] at (365,44) {$x_3$};
  \end{scope}

  % the path and the selected connection
  \draw[lightpath] (E)  -- (X1);
  \draw[lightpath] (X1) -- (X2);
  \draw[lightpath] (X2) -- (px-4-4);
  \vtx{(X1)}\vtx{(X2)}

  \node[lbl, anchor=base west] at (731,46) {$x_1$};
  \node[lbl, anchor=base west] at (496,336) {$x_2$};
  \node[lbl, anchor=base east] at (858,284) {$x_0$};
  \node[lbl, anchor=base west] at (204,284) {$x_s$};
  \node[lbl, anchor=base west] at (262,214) {$\tfrac{\Omega_k}{\omega_R}\,L_2$};
\end{tikzpicture}
    \caption{Primal Pass: reservoir selects one sensor connection}
    \label{fig:reslrb-pass1}
  \end{subfigure}
  \\[6pt]
  \begin{subfigure}[b]{\columnwidth}
    \centering
    % ResLRB, adjoint pass: the single adjoint \bar L_R at the selected pixel
% is carried back to the emitter and the path is replayed to depth R,
% applying Eq. reslrb-log-deriv at each vertex.
% Input directly by main.tex; reslrb-pass2.tex wraps it to build the PDF.
\begin{tikzpicture}[x=0.2531pt, y=0.2531pt]
  \coordinate (X1) at (726.2,61.5);
  \coordinate (X2) at (518.2,320.3);
  \coordinate (X3) at (357.2,68.3);
  \coordinate (E)  at (874.1,280.3);
  \useasboundingbox (40,0) rectangle (1000,419);

  \screen{47.6/0,74.7/1,104.1/2,136.1/3,171.3/4,209.9/5}{128.9}{32.58}{67.9}{40.7}
  \hatchpx{4}{4}
  \screenlines

  \surf{(838.1,118.4)}{(622.2,64.4)}{(739.1,10.4)}
  \surf{(517.6,367.1)}{(413,302.8)}{(604.2,280.3)}
  \begin{scope}[faded]\surf{(442.3,73.4)}{(271.3,100.4)}{(358,31)}\end{scope}
  \lrbsun[1.173]{(909.4,327)}

  % beyond the selected depth: not replayed
  \begin{scope}[faded]
    \draw[lightpath, dash pattern=on 3pt off 2pt] (X2) -- (X3);
    \vtx{(X3)}
    \node[lbl, anchor=base west] at (364,40) {$x_3$};
  \end{scope}

  \draw[lightpath] (E)  -- (X1);
  \draw[lightpath] (X1) -- (X2);
  \vtx{(X1)}\vtx{(X2)}

  % the one adjoint, carried to the start of the replay
  \draw[adjoint] (px-4-4) .. controls (232.1,289.1) and (339.7,375.6) .. (505.3,398.9)
                 .. controls (664.4,421.3) and (788.4,373) .. (849,344.5);

  \node[lbl, anchor=base west] at (731,44) {$x_1$};
  \node[lbl, anchor=base west] at (492,334) {$x_2$};
  \node[lbl, anchor=base east] at (860,276) {$x_0$};
  \node[lbl, anchor=base west] at (194,288) {$x_s$};
  \node[lbl, anchor=base west] at (268,270) {$\bar{L}_R$};
  \node[note, anchor=base west] at (765,14)
    {$\bar{L}_R\,\hat{S}\,\nabla_\theta \log f_s^{(1)}$};
  \node[note, anchor=base west] at (556,345)
    {$\bar{L}_R\,\hat{S}\,\nabla_\theta \log f_s^{(2)}$};
\end{tikzpicture}
    \caption{Adjoint Pass: replay to the selected vertex}
    \label{fig:reslrb-pass2}
  \end{subfigure}
  \caption{ResLRB two-pass structure. In the primal pass (a), candidate sensor connections are streamed into a weighted reservoir; discarded candidates are shown at reduced opacity. Only the selected connection (here at $x_2$) is splatted, reweighted by $\Omega_k / \omega_R$. In the adjoint pass (b), the image-space adjoint $\bar{L}_R$ at the selected pixel is propagated backward along the replayed path, and the log-derivative identity is applied at each vertex $x_i$.}
  \label{fig:reslrb-passes}
\end{figure}

\subsection{Reservoir Light Replay Backpropagation (ResLRB)}
\label{sec:reslrb}

Because ResLRB keeps a single sensor connection per path, its adjoint pass needs only one scalar $\bar{L}$ rather than the full set $\{\bar{L}_i\}_{i=1}^{k}$. Figure~\ref{fig:reslrb-passes} illustrates the two-pass structure.

During the primal pass the candidate connections $\{(L_i, u_i)\}_{i=1}^{k}$ are streamed into a weighted reservoir of capacity one, each carrying a scalar weight $\omega_i = \ell(L_i)$ for a scalar summary $\ell$ such as luminance.
By the standard property of weighted reservoir sampling~\cite{Vitter1985Res}, the reservoir holds at termination a single index $R \in \{1, \dots, k\}$ with
\begin{equation}
\Pr(R = i) = \frac{\omega_i}{\Omega_k},
\label{eq:reservoir-pmf}
\end{equation}
where $\Omega_k = \sum_{j=1}^{k} \omega_j$.
The primal image contribution is formed by splatting the selected connection, reweighted to compensate for the selection:
\begin{equation}
\hat{S}(\bar{x};\theta) = \frac{\Omega_k}{\omega_R} \cdot L_R(\theta),
\label{eq:reservoir-estimator}
\end{equation}
where the reweighting factor $\Omega_k / \omega_R$ is detached from $\theta$.
It is accumulated during the primal pass, outside any AD scope, and enters the adjoint pass as a constant.

The resulting gradient estimator is unbiased.
In the adjoint pass the selected index $R$ and the factor $\Omega_k / \omega_R$ are constants, so only the selected connection is differentiated, $\nabla_\theta \hat{S} = (\Omega_k / \omega_R) \, \nabla_\theta L_R$.
Averaging over the selection with the probabilities of Equation~\ref{eq:reservoir-pmf} gives
\begin{equation}
\mathbb{E}_R\!\left[\nabla_\theta \hat{S}\right]
= \sum_{i=1}^{k} \frac{\omega_i}{\Omega_k} \cdot \frac{\Omega_k}{\omega_i} \cdot \nabla_\theta L_i
= \sum_{i=1}^{k} \nabla_\theta L_i,
\label{eq:reservoir-unbiased}
\end{equation}
which is the gradient of the full path contribution $\sum_i L_i$, and the same sum without $\nabla_\theta$ shows that $\hat{S}$ is unbiased for $\sum_i L_i$.
This has to be checked directly rather than by moving the gradient inside the expectation, because the selection probabilities depend on $\theta$ through $\omega_i = \ell(L_i(\theta))$.
Differentiating the weight as well would add $-\sum_i L_i \nabla_\theta \log \Pr(R = i)$ to the expectation, which is nonzero in general.

The variance increase is proportional to the number of discarded connections. In the worst case, where every vertex produces a valid sensor connection with comparable weight, the effective sample count is reduced by a factor proportional to path length.

\subsubsection{Detached Formulation}
\label{sec:reslrb-detached}

The primal pass records the PRNG seed, the selected index $R$, and the reweighting factor $\Omega_k / \omega_R$, building no AD graph; the adjoint pass re-seeds the PRNG, replays the path to depth $R$, and enables reverse-mode AD over a local scope at each vertex.
The gradient contribution follows the log-derivative identity of Eq. \eqref{eq:prb-log-deriv}, adapted to use the reweighted splat:
\begin{equation}
\bar{L}_R \cdot \hat{S} \cdot \nabla_\theta \log f_s^{(i)},
\label{eq:reslrb-log-deriv}
\end{equation}
where $\bar{L}_R = \partial \mathcal{L} / \partial L_R$ is the image-space adjoint at the pixel receiving the selected connection.
Because only one connection survives, there is exactly one adjoint to look up, and it plays the same role as the scalar $\bar{L}$ in PRB.

\subsubsection{Attached Formulation}
\label{sec:reslrb-attached}

The attached extension adds Jacobian transport to both passes.
During the primal pass, the accumulated ray-segment Jacobian $J^{\text{ray}}_{0 \to i}$, defined in Eq. \eqref{eq:accumulated-jacobian}, is maintained alongside the reservoir.
When the reservoir accepts a new candidate at vertex $x_i$, the current $J^{\text{ray}}_{0 \to i}$ is cached with it, so that at termination the reservoir holds the accumulated Jacobian at the selected depth $R$.

After the loop, a final forward-mode differentiation step computes the sensor-connection Jacobian at the selected vertex (Figure~\ref{fig:attached-ray-uv}):
\begin{equation}
J^{\text{sensor}}_R = \frac{\partial (L_R,\, u_R)}{\partial \mathbf{p}_R} \in \mathbb{R}^{5 \times 4},
\label{eq:sensor-jacobian}
\end{equation}
which describes how a perturbation of the ray-segment parameterization $\mathbf{p}_R$ shifts both the splatted radiance $L_R$ (3 spectral channels) and the sensor coordinate $u_R$ (2 components).
Composing with the cached accumulated Jacobian gives the total splat Jacobian from the emitter parameterization to the splatted output:
\begin{equation}
J^{\text{splat}}_R = J^{\text{sensor}}_R \cdot J^{\text{ray}}_{0 \to R}.
\label{eq:splat-jacobian}
\end{equation}

\begin{figure}[t]
	\centering
	% Sensor-connection Jacobian (Figure 6): a perturbation of the segment
% p_1 = (v_0, v_1) moves the splatted radiance L_1 and the sensor
% coordinate u_1 (Eq. sensor-jacobian).
% Input directly by main.tex; ray-uv.tex wraps it to build the PDF.
\begin{tikzpicture}[x=0.2493pt, y=0.2493pt]
  \coordinate (V0) at (145.7,375.8);
  \coordinate (V1) at (343.3,112.2);
  \coordinate (U)  at (826.5,246.1);
  \useasboundingbox (10,40) rectangle (985,500);

  % area emitter with a soft glow: opaque strokes, widest and palest
  % first, matching the nine stacked 3% strokes they replace (no
  % opacity, which Texifier's typesetter drops)
  \draw[suninner!3,  line width=9pt, line join=round] (229.1,359.6) -- (60.8,289.1) -- (60.8,414.1) -- (231.1,473.2) -- cycle;
  \draw[suninner!6,  line width=8pt, line join=round] (229.1,359.6) -- (60.8,289.1) -- (60.8,414.1) -- (231.1,473.2) -- cycle;
  \draw[suninner!9,  line width=7pt, line join=round] (229.1,359.6) -- (60.8,289.1) -- (60.8,414.1) -- (231.1,473.2) -- cycle;
  \draw[suninner!12, line width=6pt, line join=round] (229.1,359.6) -- (60.8,289.1) -- (60.8,414.1) -- (231.1,473.2) -- cycle;
  \draw[suninner!14, line width=5pt, line join=round] (229.1,359.6) -- (60.8,289.1) -- (60.8,414.1) -- (231.1,473.2) -- cycle;
  \draw[suninner!17, line width=4pt, line join=round] (229.1,359.6) -- (60.8,289.1) -- (60.8,414.1) -- (231.1,473.2) -- cycle;
  \draw[suninner!20, line width=3pt, line join=round] (229.1,359.6) -- (60.8,289.1) -- (60.8,414.1) -- (231.1,473.2) -- cycle;
  \draw[suninner!22, line width=2pt, line join=round] (229.1,359.6) -- (60.8,289.1) -- (60.8,414.1) -- (231.1,473.2) -- cycle;
  \draw[suninner!25, line width=1pt, line join=round] (229.1,359.6) -- (60.8,289.1) -- (60.8,414.1) -- (231.1,473.2) -- cycle;
  \path[fill=suninner, draw=sunouter, line width=0.4pt]
    (229.1,359.6) -- (60.8,289.1) -- (60.8,414.1) -- (231.1,473.2) -- cycle;

  \surf{(206.5,194.9)}{(308,59.5)}{(500,107.9)}

  \screen{727/0,798.3/1,854.7/2,900.4/3,938.2/4,970/5}{58}{71.6}{155.8}{43.05}
  \fillpx{1}{2}
  \screenlines

  % local frames
  \draw[frameu] (V0) -- ++(0,-40);  \draw[framev] (V0) -- ++(-40,0);
  \draw[frameu] (V1) -- ++(0,-40);  \draw[framev] (V1) -- ++(-40,0);
  \draw[frameu, -{Stealth[length=2pt, width=1.8pt]}] (U) -- ++(0,22);
  \draw[framev, -{Stealth[length=2pt, width=1.8pt]}] (U) -- ++(22,0);

  \draw[ray, shorten >=2.6pt] (V0) -- (V1);
  \draw[ray, shorten >=1.2pt] (V1) -- (U);
  \draw[jac] (238,336) .. controls (338.3,243.1) and (541.5,230.2) .. (685.8,250);
  \fill[vertexorange] (V0) circle (2.4pt);
  \fill[vertexorange] (V1) circle (2.4pt);
  \fill[vertexorange] (U)  circle (1.3pt);

  \node[lbl, anchor=base west] at (156,382) {$\mathbf{v}_0$};
  \node[lbl, anchor=base west] at (354,84)  {$\mathbf{v}_1$};
  \node[lbl, anchor=base east] at (790,250) {$u_1$};
  \node[lbl, anchor=base west] at (390,280) {$J^{\text{sensor}}_1$};
\end{tikzpicture}
	\caption{Sensor-connection Jacobian $J^{\text{sensor}}_i$. A perturbation of the ray-segment parameterization $\mathbf{p}_i$ shifts both the splatted radiance $L_i$ and the sensor coordinate $u_i$. The composition $J^{\text{splat}}_i = J^{\text{sensor}}_i \cdot J^{\text{ray}}_{0 \to i}$ maps perturbations at the path origin to shifts in the splatted output.}
 	\label{fig:attached-ray-uv}
\end{figure}

During the adjoint pass, $J^{\text{ray}}_{0 \to j}$ is rebuilt at each replayed vertex by the same forward-mode process used in the primal pass.
At vertex $x_j$, the adjoint is projected into the local frame via the inverse, as in Eq. \eqref{eq:jacobian-projection}.
Two backward operations are performed: one through the radiance channel weighted by $\bar{L}_R$, and one through the position channel weighted by the position adjoint $\partial \mathcal{L} / \partial u_R$.
The computation of this position adjoint is described in Section~\ref{sec:uv-adjoint}.

Because the reservoir selects a single vertex, the attached formulation adds only constant storage per sample: one $4 \times 4$ matrix for $J^{\text{ray}}_{0 \to R}$ and one two-component vector for the position adjoint.

\subsection{LRB-3-Pass}
\label{sec:lrb-3pass}

LRB-3-pass retains all $k$ sensor connections rather than sampling one, at the cost of a third path traversal. The core observation is that the downstream sum $\sum_{i > j} \bar{L}_i \cdot L_i$ needed at each vertex during the backward pass can be precomputed in a forward scan and then consumed by subtraction, mirroring the remaining-radiance pattern of PRB but extended from a single contribution to a sum over sensor connections. Figure~\ref{fig:lrb-passes} illustrates the three-pass structure.

\begin{figure}[t]
  \centering
  % LRB-3-pass in one figure: three passes over the same light path,
% stacked, with what each pass hands to the next drawn between rows.
% Input directly by main.tex; lrb-3pass.tex wraps it to build the PDF.
\input{lrb-scene}
\begin{tikzpicture}[x=0.26557pt, y=0.26557pt]
  \def\rowgap{430}
  \useasboundingbox (60,{-2*\rowgap-26}) rectangle (975,333);

  % ---- Pass 1: primal -------------------------------------------
  \begin{scope}
    \lrbscene
    \draw[conn, shorten <=3pt] (X2) -- (px-4-4);
    \draw[conn, shorten <=3pt] (X1) -- (px-1-3);
    \draw[conn, shorten <=3pt] (X3) -- (px-3-1);
    \lightpaths
    \vertices
    \node[lbl, anchor=base west] at (300,246) {$L_2$};
    \node[lbl, anchor=base west] at (373,140) {$L_1$};
    \node[lbl, anchor=base west] at (245,100) {$L_3$};
    \passtitle{(a) Primal}
  \end{scope}

  % ---- Pass 2: accumulate ---------------------------------------
  \begin{scope}[yshift=-\rowgap*0.26557pt]
    \lrbscene
    \adjointreads
    \lightpaths
    \vertices
    \node[note, anchor=base west] at (762,36)  {$+\,\bar{L}_1 L_1$};
    \node[note, anchor=base west] at (590,282) {$+\,\bar{L}_2 L_2$};
    \node[note, anchor=base west] at (420,-14) {$+\,\bar{L}_3 L_3$};
    \passtitle{(b) Accumulate}
  \end{scope}

  % ---- Pass 3: backward -----------------------------------------
  \begin{scope}[yshift=-2*\rowgap*0.26557pt]
    \lrbscene
    \adjointreads
    \lightpaths
    \vertices
    \node[lbl, anchor=base west] at (795,150) {$\delta_{\text{total}}$};
    \node[lbl, anchor=base west] at (634,158) {$\delta_{\text{rem}}$};
    \node[lbl, anchor=base west] at (470,168) {$\delta_{\text{rem}}$};
    \node[note, anchor=base west] at (735,-16)
      {$\delta_{\text{rem}}\,\nabla_\theta \log f_s^{(1)}$};
    \node[note, anchor=base west] at (562,284)
      {$\delta_{\text{rem}}\,\nabla_\theta \log f_s^{(2)}$};
    \node[note, anchor=base west] at (420,-16)
      {$\delta_{\text{rem}}\,\nabla_\theta \log f_s^{(3)}$};
    \passtitle{(c) Backward}
  \end{scope}

  % ---- hand-offs between passes ---------------------------------
  \tikzset{handoff/.style={draw=black!70, line width=0.7pt,
           -{Stealth[length=4pt, width=3.4pt]}}}
  \draw[handoff] (100,62) -- (100,{-\rowgap+300});
  \node[note, anchor=west, align=left] at (112,{-\rowgap/2+150})
    {$\delta I$: image-space adjoint};
  \draw[handoff] (100,{-\rowgap+62}) -- (100,{-2*\rowgap+300});
  \node[note, anchor=west] at (112,{-3*\rowgap/2+150})
    {$\delta_{\text{total}} = \sum_{i} \bar{L}_i L_i$};
\end{tikzpicture}
  \caption{LRB-3-pass structure. The primal pass (a) traces the light path and splats contributions to the sensor. The accumulation pass (b) re-traces the path, reading per-vertex adjoints from the image gradient tensor and summing $\bar{L}_i \cdot L_i$ along the path. The backward pass (c) re-traces once more, consuming the accumulated sum by subtraction and backpropagating at each vertex. Gray arrows mark what each pass hands to the next. Only $\delta_{\text{rem}}$ and the local AD scope are held in memory at any point.}
  \label{fig:lrb-passes}
\end{figure}

\subsubsection{Detached Formulation}
\label{sec:lrb3-detached}

The algorithm proceeds in three passes over the same path, reconstructed each time from the same PRNG seed.

\textbf{Pass 1 (Primal).}
Light paths are traced and their contributions splatted to an image block as in a standard light tracer.
The image is developed, the loss $\mathcal{L}$ evaluated, and the gradient $\partial \mathcal{L} / \partial I[p]$ backpropagated through the film pipeline to produce a per-pixel adjoint tensor $\delta I$, stored as a read-only lookup for subsequent passes.

\textbf{Pass 2 (Accumulate).}
Each path is re-traced from the same seed.
At every vertex $x_i$ with a valid sensor connection, the image-space adjoint $\bar{L}_i$ is read from $\delta I$ at sensor coordinate $u_i$ using a filter-weighted lookup matching the reconstruction filter used during splatting.
The product $\bar{L}_i \cdot L_i$ is accumulated into a running per-sample sum:
\begin{equation}
\delta_{\text{total}} = \sum_{i=1}^{k} \bar{L}_i \cdot L_i.
\label{eq:delta-total}
\end{equation}
The accumulator is the sum of element-wise products $\bar{L}_i \cdot L_i$.

\textbf{Pass 3 (Backward).}
The path is re-traced a third time with $\delta_{\text{rem}}$ initialized to $\delta_{\text{total}}$.
At each vertex $x_i$, two backward operations are performed.
First, the per-vertex adjoint $\bar{L}_i$ is read from $\delta I$ and gradients are backpropagated through the local splat computation:
\begin{equation}
\nabla_\theta f_{\text{splat}}^{(i)} \cdot \bar{L}_i.
\label{eq:3pass-splat-grad}
\end{equation}
Second, the current vertex's contribution is subtracted,
\begin{equation}
\delta_{\text{rem}} \leftarrow \delta_{\text{rem}} - \bar{L}_i \cdot L_i,
\label{eq:delta-subtract}
\end{equation}
leaving $\delta_{\text{rem}} = \sum_{j > i} \bar{L}_j \cdot L_j$, the downstream adjoint.
The log-derivative identity is applied using $\delta_{\text{rem}}$ as the weight:
\begin{equation}
\delta_{\text{rem}} \cdot \nabla_\theta \log f_s^{(i)}.
\label{eq:3pass-throughput-grad}
\end{equation}
After both operations, the AD graph for vertex $x_i$ is discarded.
The live state at any point consists of $\delta_{\text{rem}}$ (one spectral value per sample), the path throughput, and the local AD scope, all constant in path length.

\subsubsection{Attached Formulation}
\label{sec:lrb3-attached}

The attached extension adds a second accumulator $A_{\text{pos}}$ alongside $\delta_{\text{total}}$ to handle the position adjoint channel.

In Pass~2, at each vertex $x_i$ with a valid sensor connection, the position adjoint $\partial \mathcal{L} / \partial u_i$ is computed by the procedure described in Section~\ref{sec:uv-adjoint}.
The accumulated Jacobian $J^{\text{ray}}_{0 \to i}$ and the sensor-connection Jacobian $J^{\text{sensor}}_i$ (defined in Eq. \eqref{eq:sensor-jacobian}) are rebuilt on the fly, and their composition gives the per-vertex splat Jacobian $J^{\text{splat}}_i = J^{\text{sensor}}_i \cdot J^{\text{ray}}_{0 \to i}$ as in Eq. \eqref{eq:splat-jacobian}.
The position adjoint is projected through this Jacobian and accumulated:
\begin{equation}
A_{\text{pos}} = \sum_{i=1}^{k} \begin{pmatrix} 0 & 0 & \frac{\partial \mathcal{L}}{\partial u_{i,x}} & \frac{\partial \mathcal{L}}{\partial u_{i,y}} \end{pmatrix} J^{\text{splat}}_i.
\label{eq:a-pos-accum}
\end{equation}
This produces a four-component vector per sample.

In Pass~3, $A_{\text{pos}}$ is consumed by subtraction alongside $\delta_{\text{rem}}$.
At vertex $x_j$, the remainder is projected into the local frame:
\begin{equation}
A_{\text{pos,local}} = A_{\text{pos,rem}} \cdot \left(J^{\text{ray}}_{0 \to j}\right)^{-1},
\label{eq:a-pos-project}
\end{equation}
where $J^{\text{ray}}_{0 \to j}$ is rebuilt during replay.
The last two components of $A_{\text{pos,local}}$ give the position adjoint in the local parameterization at $x_j$, which is backpropagated through the two-point parameterization alongside the radiance adjoint from Eq. \eqref{eq:3pass-throughput-grad}.
The total per-sample storage for the attached case is $\delta_{\text{rem}}$ (one spectral value) plus $A_{\text{pos}}$ (four floats), independent of path length.

\subsection{The Splat Position Adjoint}
\label{sec:uv-adjoint}

Both attached formulations require the position adjoint $\partial \mathcal{L} / \partial u_i$ at each vertex with a valid sensor connection. The splatting operation deposits radiance $L_i$ into the image through a reconstruction filter $F$ centered at $u_i$:
\begin{equation}
I[p] \mathrel{+}= F(u_i, p) \cdot L_i \cdot s,
\label{eq:splat-filter}
\end{equation}
where $s$ is a per-sample normalization factor and the sum runs over pixels $p$ in the filter's support.
The gradient of the loss with respect to $u_i$ involves the spatial derivative of $F$:
\begin{equation}
\frac{\partial \mathcal{L}}{\partial u_i} = \sum_{p} \frac{\partial \mathcal{L}}{\partial I[p]} \cdot L_i \cdot s \cdot \frac{\partial F(u_i, p)}{\partial u_i}.
\label{eq:dpos-analytic}
\end{equation}

We compute $\partial \mathcal{L} / \partial u_i$ by reverse-mode differentiation through the filter weighted lookup.
Within a local AD scope, we enable gradients on the position $u_i$, perform a filter-weighted read from the adjoint tensor $\delta I$, multiply by $L_i$, and backpropagate to obtain $\partial \mathcal{L} / \partial u_i$.
This is a small self-contained computation involving only the filter kernel and a gather from $\delta I$.
In ResLRB it is computed once for the selected vertex; in LRB-3-pass it is computed at every vertex during Pass~2 and projected through $J^{\text{splat}}_i$ before accumulation into $A_{\text{pos}}$.

\section{Implementation}
\label{sec:implementation}

\begin{figure}[b]
\begin{mdframed}[style=nicebox]
\begin{lstlisting}[style=nicestyle]
def lrb_3pass_backward(seed, grad_in):
  # Pass 1: primal
  trace_light_path(seed)
  $\delta$I = backprop_film(grad_in)

  # Pass 2: accumulate
  $\delta_{\text{total}}$ = 0
  for i in range(max_depth):
    v = intersect(ray)
    L, u = connect_to_sensor(v, $\beta$)
    if L is valid:
      $\bar{L}_i$ = read_adjoint($\delta$I, u)
      $\delta_{\text{total}}$ += $\bar{L}_i$ * L
    advance_path(v)

  # Pass 3: backward
  $\delta_{\text{rem}}$ = $\delta_{\text{total}}$
  for i in range(max_depth):
    v = intersect(ray)
    L, u = connect_to_sensor(v, $\beta$)
    if L is valid:
      $\bar{L}_i$ = read_adjoint($\delta$I, u)
      backward_grad(L, $\bar{L}_i$)
      $\delta_{\text{rem}}$ -= $\bar{L}_i$ * L
    backward_grad(f_s, $\delta_{\text{rem}}$ / f_s)
    advance_path(v)
\end{lstlisting}
\end{mdframed}
\captionof{figure}{LRB-3-pass backward structure. Pass~2 accumulates the adjoint-weighted radiance; Pass~3 consumes it by subtraction while backpropagating at each vertex.}
\label{lst:lrb3-backward}
\end{figure}

We implement all methods as custom integrators in Mitsuba~3~\cite{Mitsuba3} on top of DrJit~\cite{Jakob2020DrJit}.
The tracing loop runs in symbolic mode, which records the loop body as a single fused kernel and executes all iterations in one launch, preserving the constant-memory property.

The inverse of the accumulated Jacobian $J^{\text{ray}}_{0 \to j}$ can be singular at vertices involving diffuse scattering or near-grazing intersections; we substitute the zero matrix when $|\det J^{\text{ray}}_{0 \to j}|$ falls below $10^{-8}$ or its Frobenius norm exceeds $10^{6}$.
The norm test matters on long specular chains, where reflection keeps the determinant near one while the singular values spread and the single-precision inverse overflows (supplementary Section~S4).
As a complementary measure, we inject zero-mean stochastic noise into the local Jacobian at non-specular vertices, following the regularization scheme of \citet{Vicini2021PathReplay}.

\textbf{ResLRB.}
Both passes follow the pseudocode of \citet{Beresna2026Stochastic}.

\textbf{LRB-3-pass.}
Listing~\ref{lst:lrb3-backward} shows the three-pass backward procedure.
The adjoint tensor $\delta I$ is constructed between Pass~1 and Pass~2 by enabling gradients on the image block's internal tensor, developing the image, and backpropagating the loss gradient through the film pipeline.
The result is wrapped in a read-only image block configured with the same reconstruction filter as the primal block, so that a read at position $u_i$ returns the filter-matched adjoint.

\textbf{Attached extensions.}
Listing~\ref{lst:read-adjoint} reads the radiance and position adjoints from $\delta I$ in a single local AD scope, and Listing~\ref{lst:lrb3-attached-pass2} shows the attached accumulation pass of LRB-3-pass.

\begin{figure}[t]
\begin{mdframed}[style=nicebox]
\begin{lstlisting}[style=nicestyle]
def read_adjoint($\delta$I, u, L):
  u_diff = detach_copy(u)
  enable_grad(u_diff)
  $\bar{L}$ = filter_read($\delta$I, u_diff)
  backward_from(dot($\bar{L}$, detach(L)))
  $\delta$pos = grad(u_diff)
  return $\bar{L}$, $\delta$pos
\end{lstlisting}
\end{mdframed}
\captionof{figure}{Joint radiance and position adjoint read. The filter-weighted lookup is differentiated with respect to the splat position in a local AD scope, yielding both $\bar{L}_i$ and $\partial \mathcal{L} / \partial u_i$.}
\label{lst:read-adjoint}
\end{figure}

\begin{figure}[t]
\begin{mdframed}[style=nicebox]
\begin{lstlisting}[style=nicestyle]
# LRB-3-pass attached: Pass 2
$\delta_{\text{total}}$ = 0, $A_{\text{pos}}$ = 0
$J^{\text{ray}}_{0 \to i}$ = identity()
for i in range(max_depth):
  v = intersect(ray)
  L, u = connect_to_sensor(v, $\beta$)
  if L is valid:
    $\bar{L}_i$, $\delta$pos = read_adjoint($\delta$I, u, L)
    $\delta_{\text{total}}$ += $\bar{L}_i$ * L
    $J^{\text{sensor}}_i$ = forward_jacobian(v, u)
    $J^{\text{splat}}_i$ = $J^{\text{sensor}}_i$ * $J^{\text{ray}}_{0 \to i}$
    $A_{\text{pos}}$ += [0, 0, $\delta$pos.x, $\delta$pos.y] * $J^{\text{splat}}_i$
  $J^{\text{ray}}_i$ = forward_jacobian(v)
  $J^{\text{ray}}_{0 \to i}$ = $J^{\text{ray}}_i$ * $J^{\text{ray}}_{0 \to i}$
  advance_path(v)
\end{lstlisting}
\end{mdframed}
\captionof{figure}{LRB-3-pass attached, Pass~2. The detached accumulation of $\delta_{\text{total}}$ is augmented with forward-mode Jacobian transport and position adjoint accumulation into $A_{\text{pos}}$. The \texttt{read\_adjoint} call (Listing~\ref{lst:read-adjoint}) returns both $\bar{L}_i$ and $\partial \mathcal{L} / \partial u_i$.}
\label{lst:lrb3-attached-pass2}
\end{figure}
 
 \begin{figure*}[t]
    \centering
    \setlength{\tabcolsep}{1.2pt}
    \renewcommand{\arraystretch}{1.0}
    \newcolumntype{C}{>{\centering\arraybackslash}m{0.18\textwidth}}
    \newlength{\gradh}\setlength{\gradh}{0.10125\textwidth}
    \begin{tabularx}{\textwidth}{@{} >{\centering\arraybackslash}m{0.03\textwidth} *{5}{C} @{}}
        & \textbf{Scene}
        & \texttt{ptracer}
        & \texttt{lrb\_2pass}
        & \texttt{reslrb}
        & \texttt{lrb\_3pass} \\ \addlinespace[4pt]

        \rotatebox[origin=c]{90}{\textbf{Room}} &
        \includegraphics[height=\gradh,keepaspectratio]{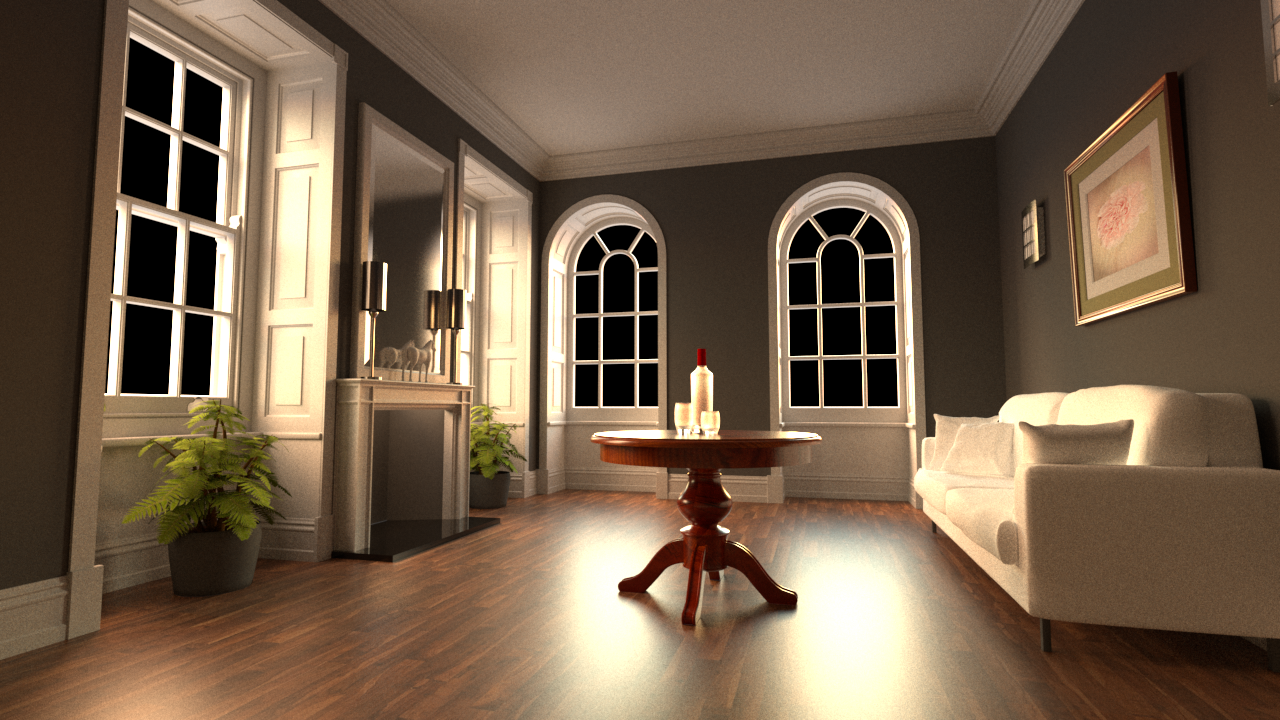} &
        \includegraphics[width=\linewidth,height=\gradh]{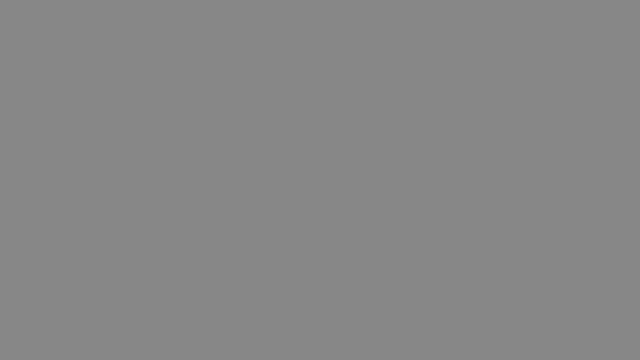} &
        \includegraphics[width=\linewidth,height=\gradh]{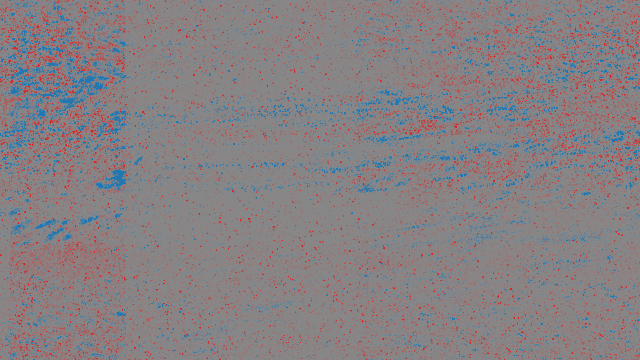} &
        \includegraphics[width=\linewidth,height=\gradh]{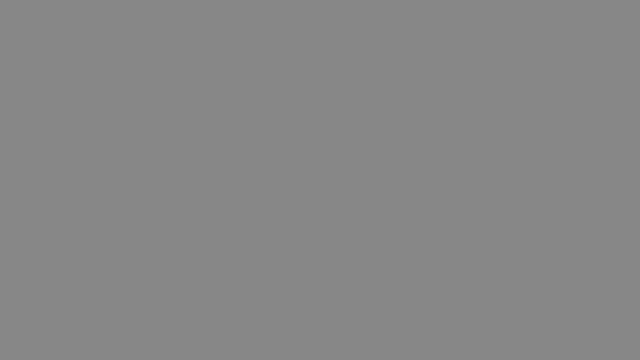} &
        \includegraphics[width=\linewidth,height=\gradh]{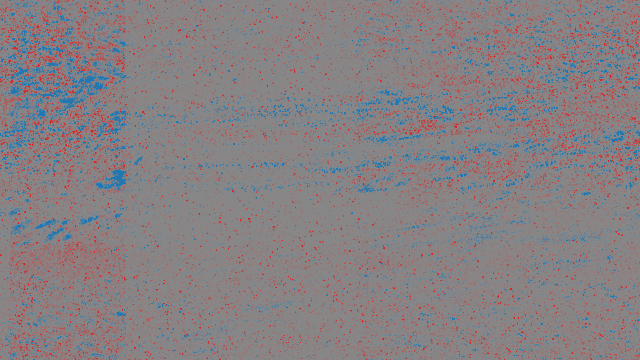} \\ \addlinespace[3pt]

        \rotatebox[origin=c]{90}{\textbf{Slab}} &
        \includegraphics[height=\gradh,keepaspectratio]{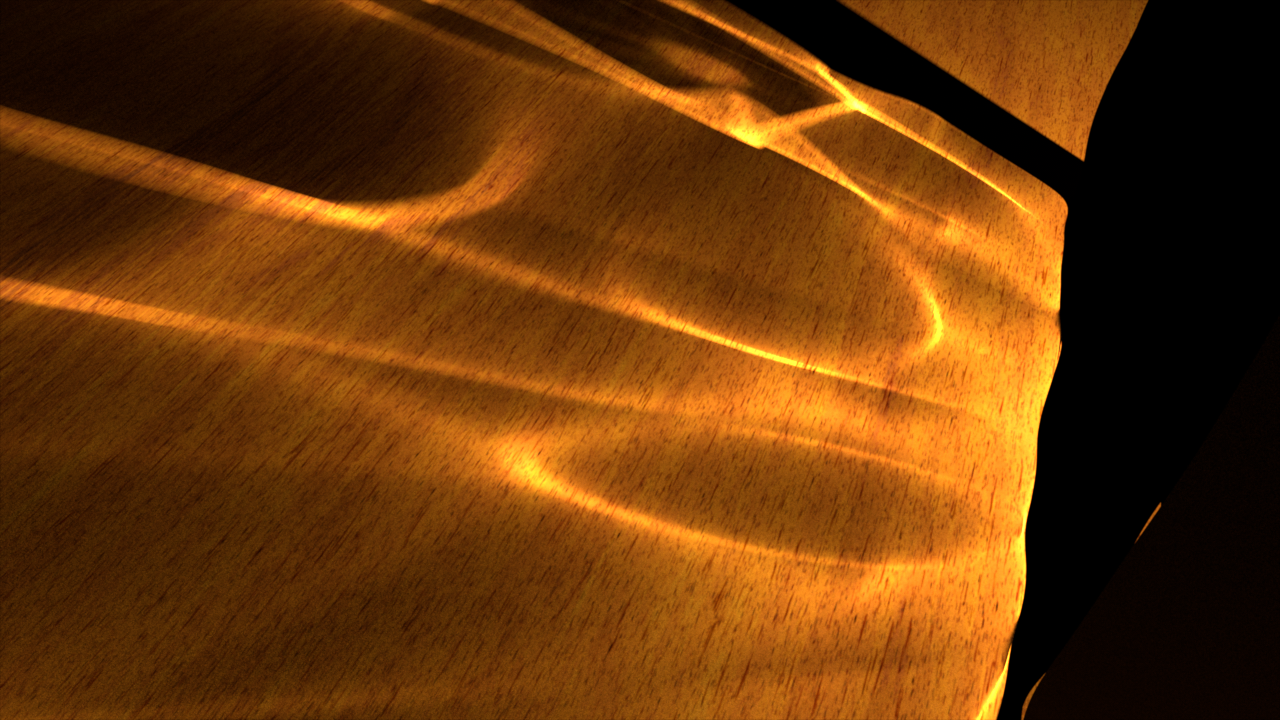} &
        \includegraphics[width=\linewidth,height=\gradh]{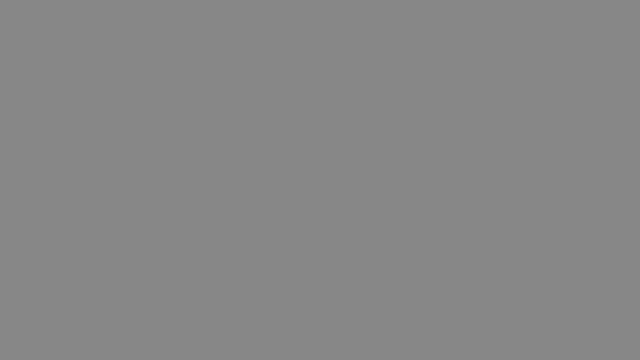} &
        \includegraphics[width=\linewidth,height=\gradh]{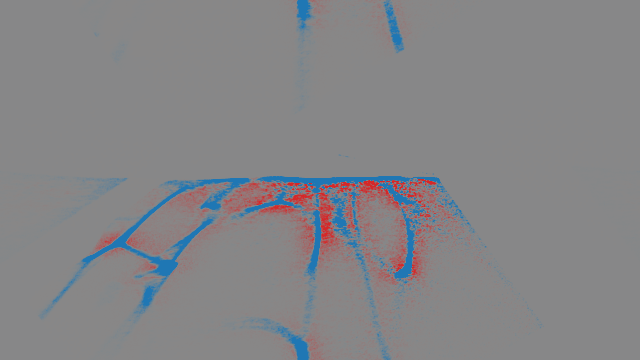} &
        \includegraphics[width=\linewidth,height=\gradh]{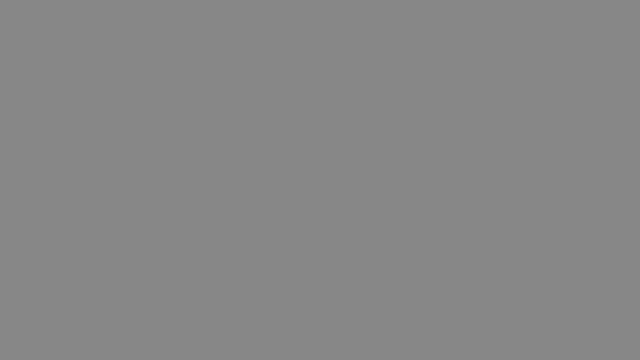} &
        \includegraphics[width=\linewidth,height=\gradh]{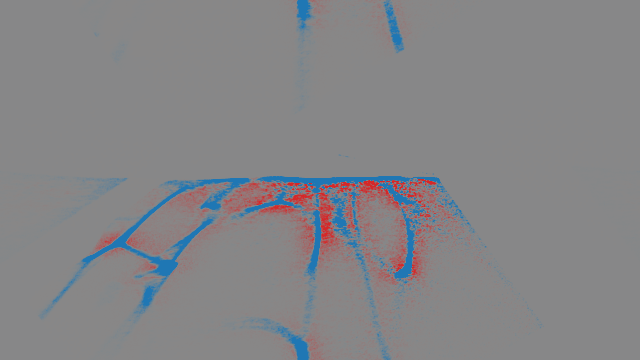} \\ \addlinespace[3pt]

        \rotatebox[origin=c]{90}{\textbf{Mirror}} &
        \includegraphics[height=\gradh,keepaspectratio]{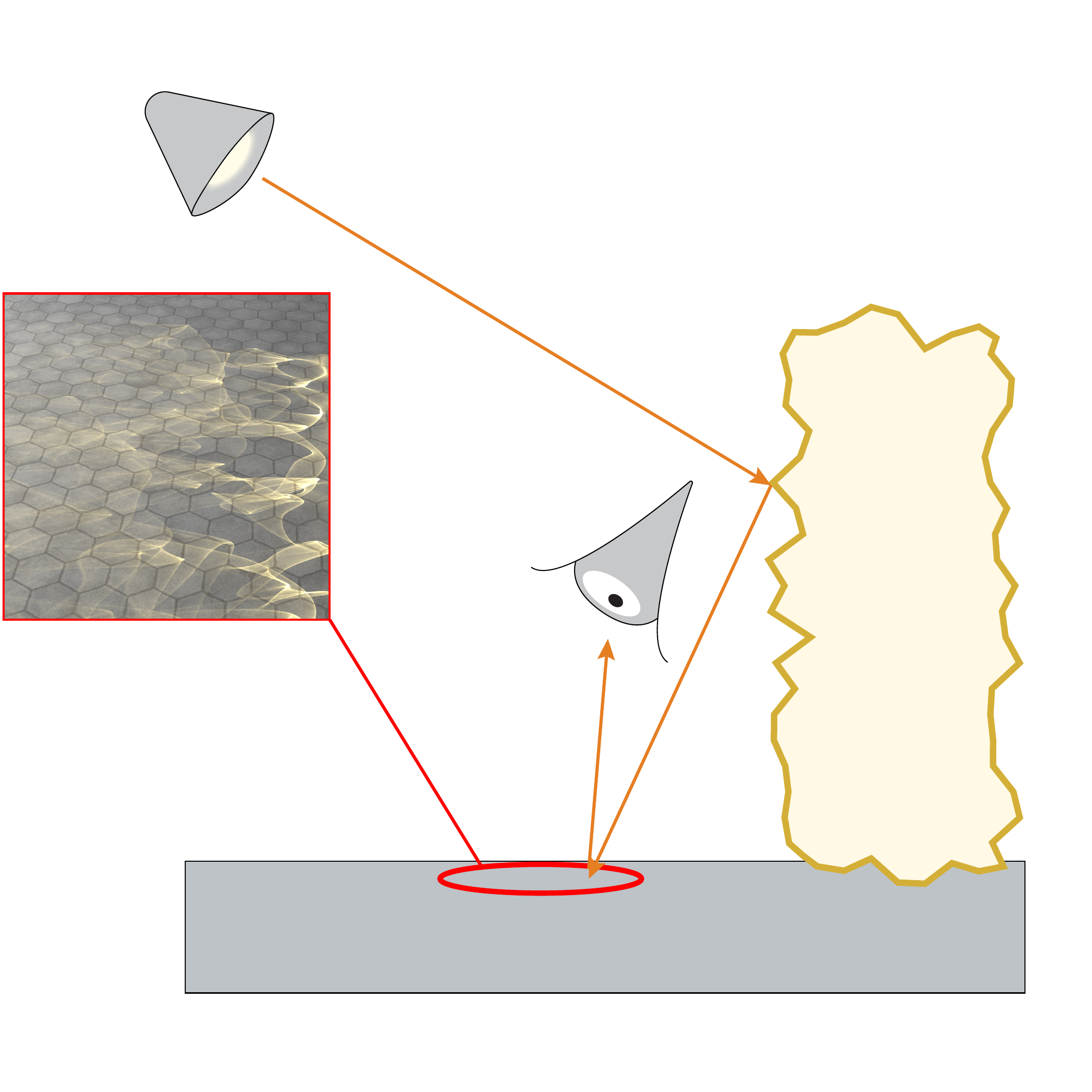} &
        \includegraphics[width=\linewidth,height=\gradh]{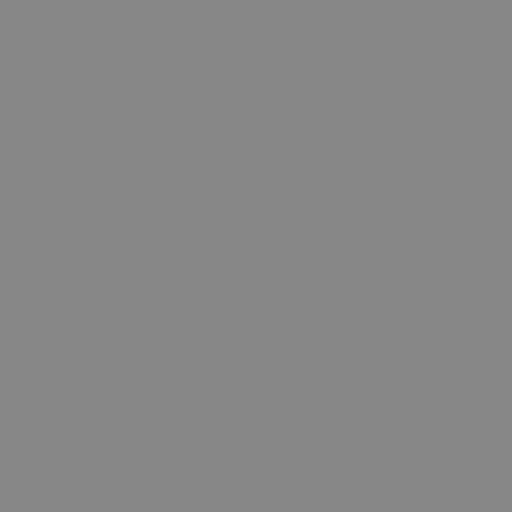} &
        \includegraphics[width=\linewidth,height=\gradh]{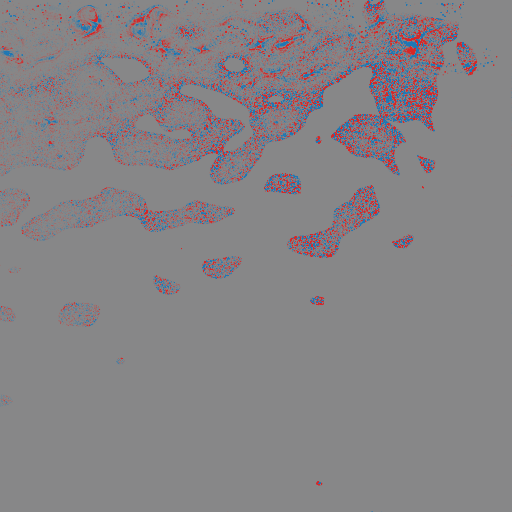} &
        \includegraphics[width=\linewidth,height=\gradh]{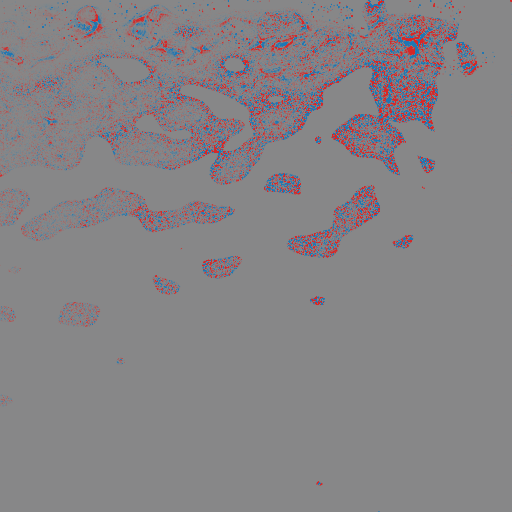} &
        \includegraphics[width=\linewidth,height=\gradh]{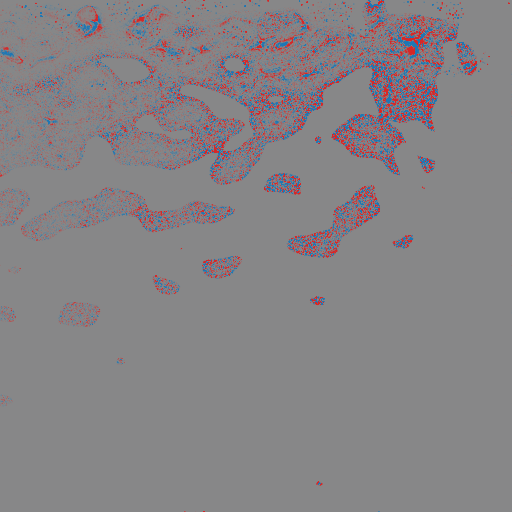} \\ \addlinespace[3pt]

        \rotatebox[origin=c]{90}{\textbf{Glass}} &
        \includegraphics[height=\gradh,keepaspectratio]{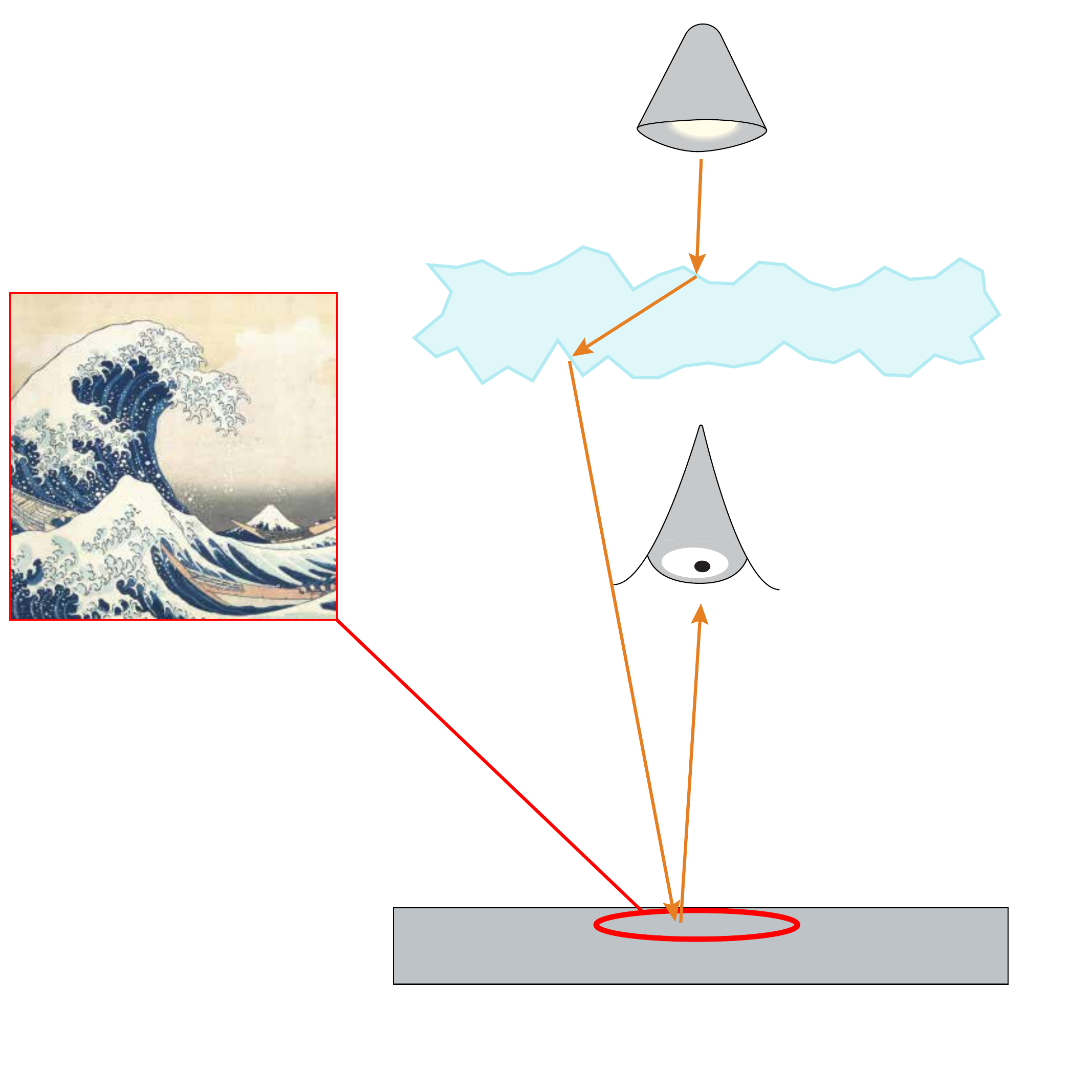} &
        \includegraphics[width=\linewidth,height=\gradh]{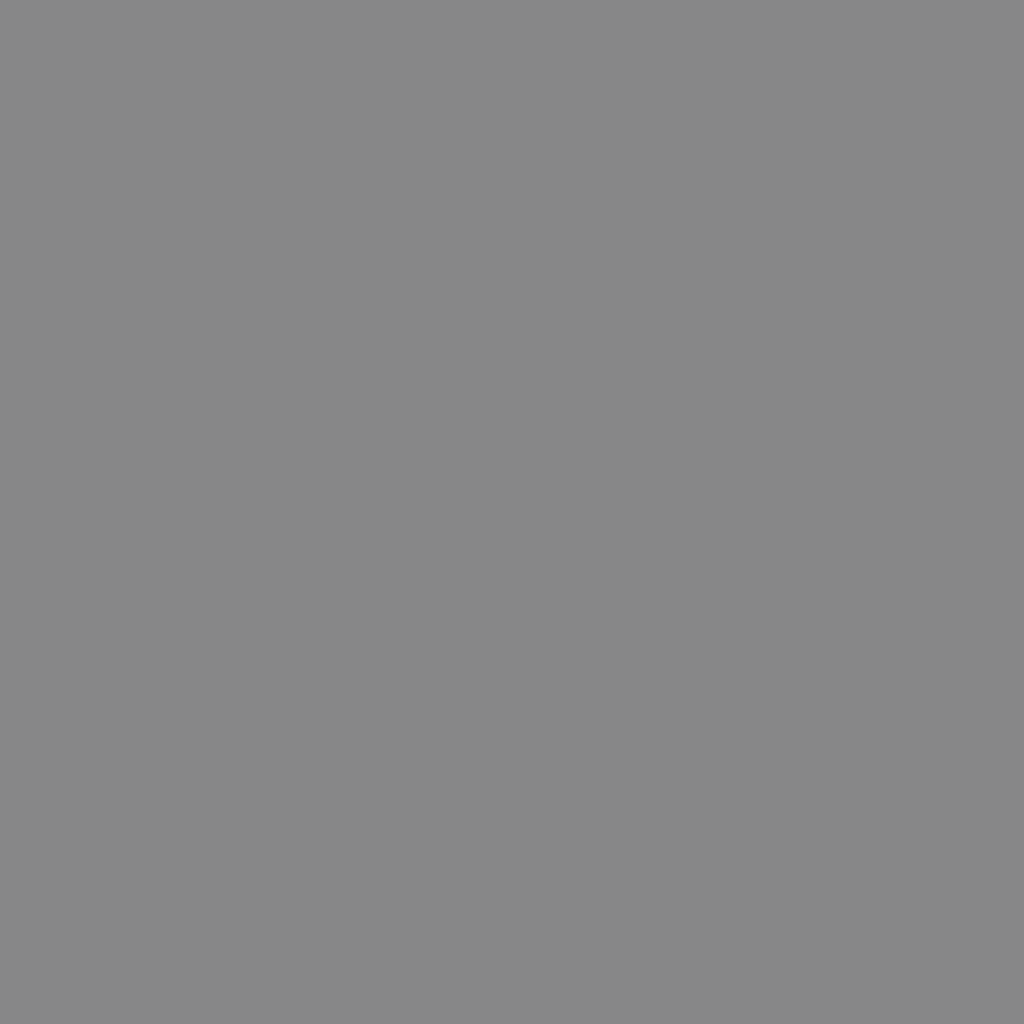} &
        \includegraphics[width=\linewidth,height=\gradh]{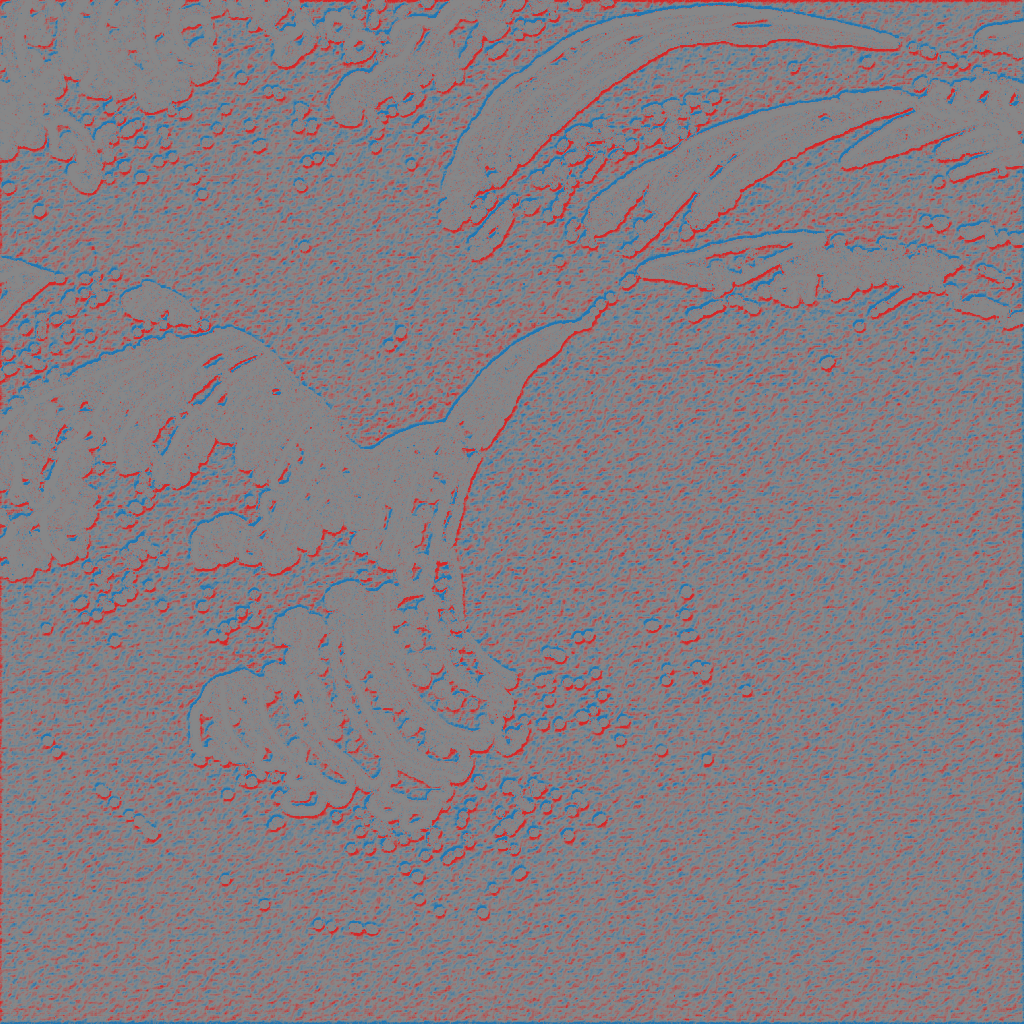} &
        \includegraphics[width=\linewidth,height=\gradh]{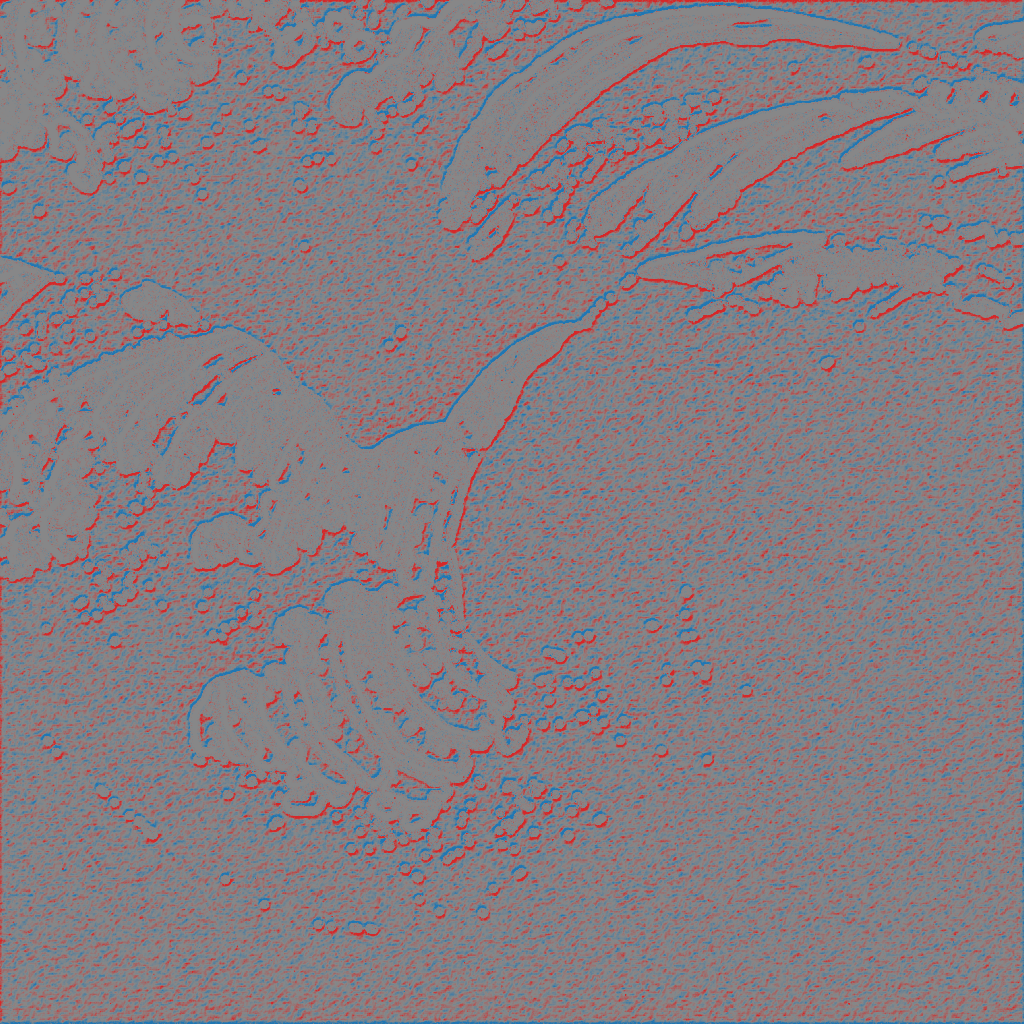} &
        \includegraphics[width=\linewidth,height=\gradh]{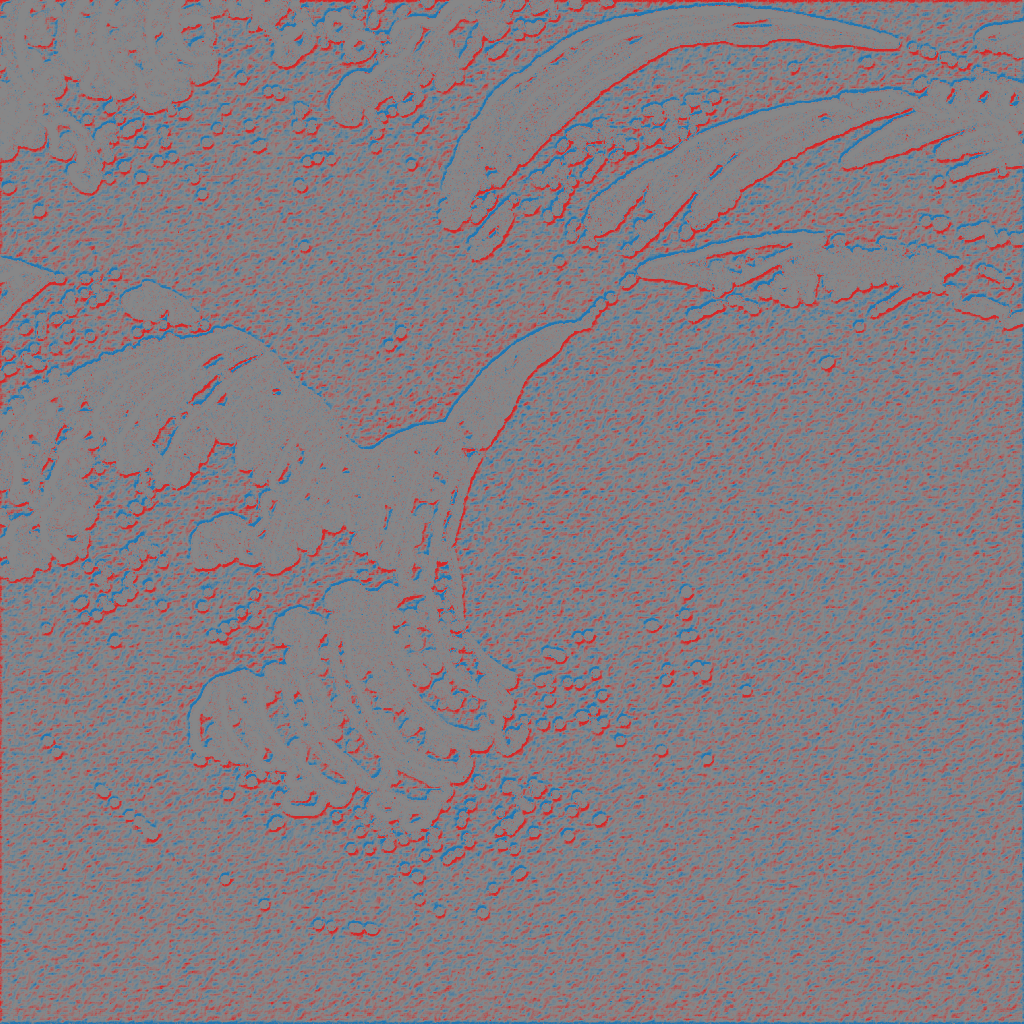} \\
    \end{tabularx}

    \caption{Gradient validation against a blurred-render target: detached above, attached below. Gradients are shown on one channel with a symmetric red--blue colormap. Every variant matches \texttt{ptracer} up to Monte Carlo noise.}
    \label{fig:gradient-validation}
\end{figure*}
 
\section{Results}
\label{sec:results}

This section validates the gradients of all four methods, measures how their memory and time scale with path length, and applies them to two optimizations, a fog and a visibility discontinuity; the light-guide recovery of Figure~\ref{fig:teaser} is detailed in the supplementary material.
All experiments are performed on a workstation with an NVIDIA GeForce RTX 5080 GPU (16\,GB VRAM) and an AMD Ryzen 9 9950X CPU, using the CUDA/OptiX \cite{Nickolls2008CUDA, Parker2010Optix} backend.
We evaluate seven integrator variants spanning the design space: naive AD light tracing (\texttt{ptracer}), LRB-2-pass (\texttt{lrb\_2pass}, detached and attached), which buffers the per-vertex data, ResLRB (\texttt{reslrb}, detached and attached), and LRB-3-pass (\texttt{lrb\_3pass}, detached and attached).

\textbf{Scenes.} We use five scenes across the experiments: \textbf{Living Room} (open diffuse interior, high path--sensor connectivity, referred to as Room in plots), \textbf{Glass Slab} (dielectric slab over a diffuse receiver), \textbf{Mirror} (roughened conductor producing a caustic on a diffuse receiver), \textbf{Glass Panel} (flat dielectric panel with a normal map caustic, referred to as Glass in plots), and \textbf{Guide} (a straight dielectric light guide carrying an image to a diffuse receiver).
Living Room and Glass Slab are used for detached validation, Mirror and Glass Panel for attached validation, and Guide for the path-length sweep.

\begin{figure}[t]
  \centering
  \includegraphics[width=\linewidth]{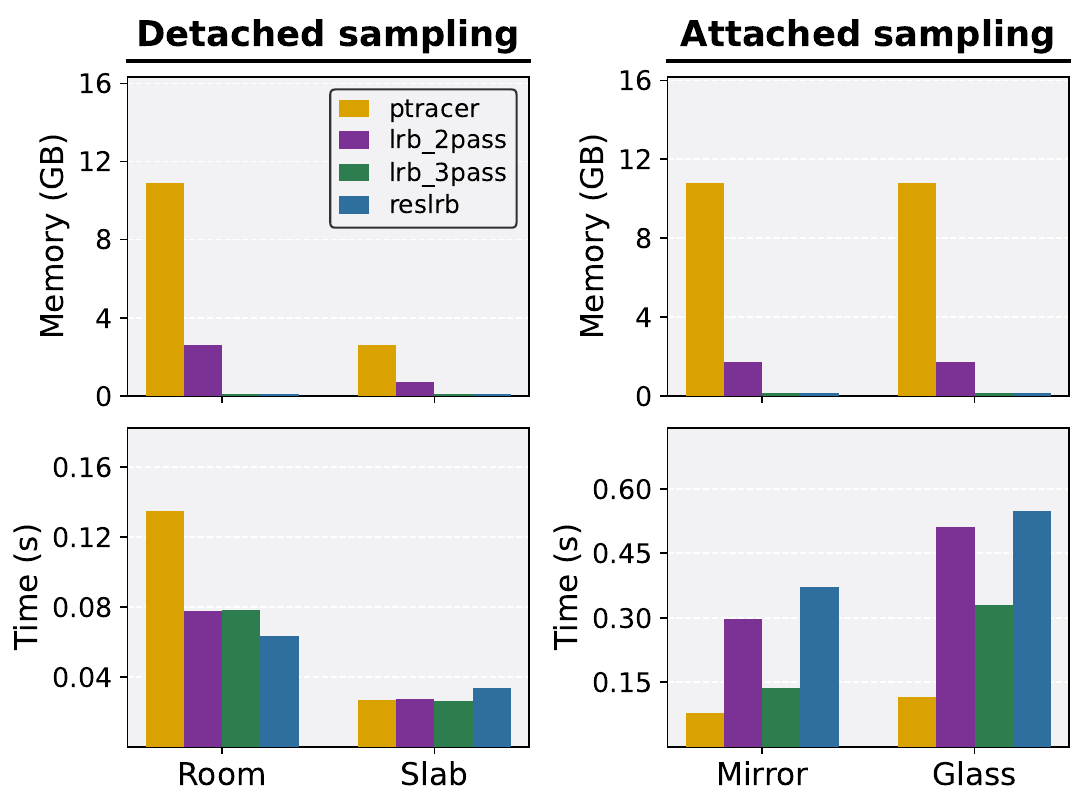}
  \caption{Memory and per-iteration time across the validation scenes.}
  \label{fig:rendering-stats-2x2}
\end{figure}

\begin{figure*}[t]
  \centering
  \includegraphics[width=\textwidth]{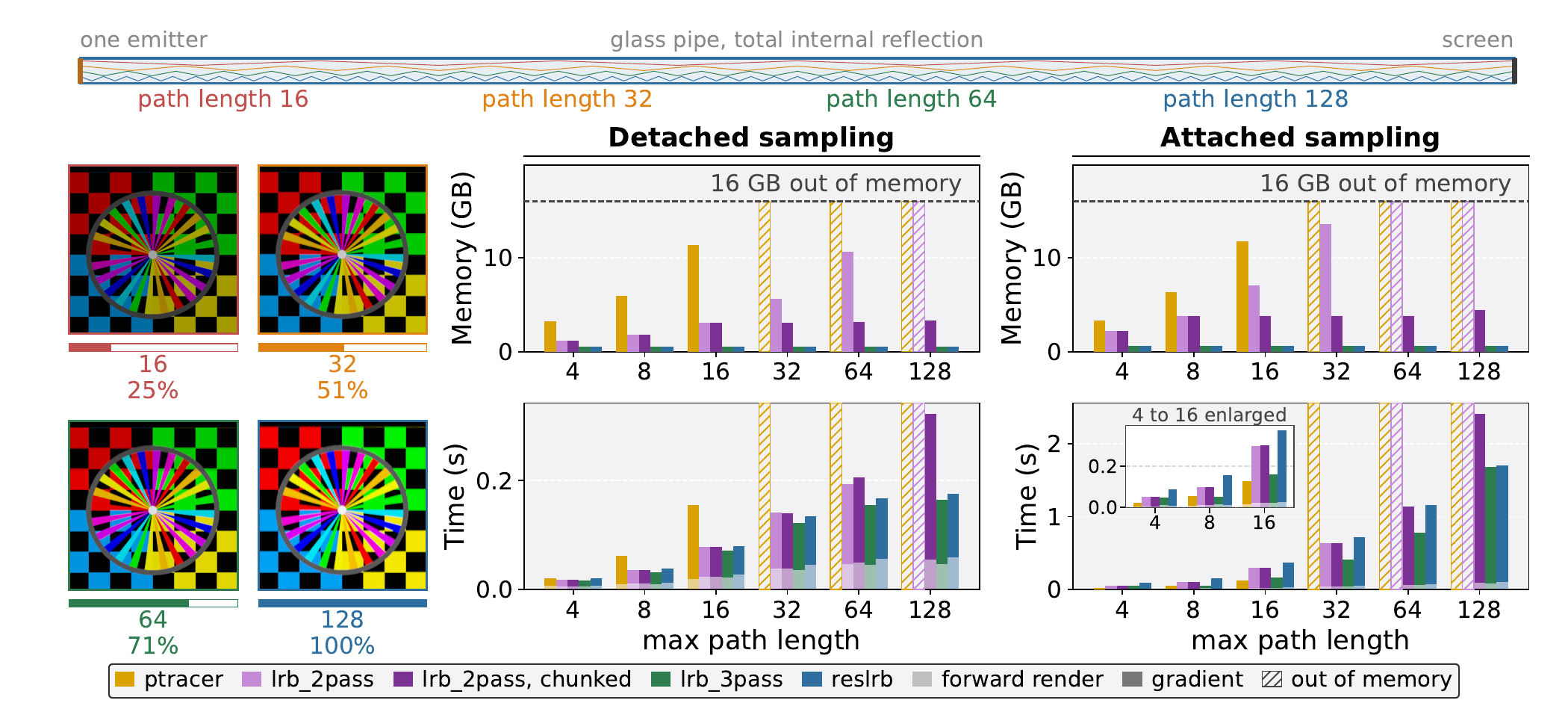}
  \caption{Path-length sweep on the Guide scene (320$^2$, 64~spp). A glass pipe carries a picture to a screen by total internal reflection along four directions, which arrive after 12, 28, 60 and 124 reflections; each screen is labeled with its share of the light that arrives at path length 128. Attached differentiates the wall normal map, detached the wall roughness; both target a blurred render. Bars are split at the forward render, and hatched bars did not fit in 16\,GB. LRB-2-pass is shown with and without chunking.}
  \label{fig:depth-sweep}
\end{figure*}

\subsection{Gradient Correctness}
\label{sec:results-gradient}
We validate gradients by backpropagating an $L_2$ loss through the rendered image and comparing the resulting parameter gradient.

\subsubsection{Detached Validation}
\label{sec:detached-validation}

We validate detached gradients against the \texttt{ptracer} reference on two scenes (Figure~\ref{fig:gradient-validation}), using a synthetic target constructed by blurring the primal render with a Gaussian kernel.
For the Living Room scene (path length 16, 32~spp) we differentiate the diffuse reflectance of the floor; for the Glass Slab (path length 4, 32~spp) we differentiate the diffuse reflectance of the receiving plane.
In both scenes \texttt{reslrb}, \texttt{lrb\_2pass}, and \texttt{lrb\_3pass} agree with the reference up to Monte Carlo noise, with \texttt{reslrb} exhibiting more noise for the Living Room scene.

\subsubsection{Attached Validation}
\label{sec:attached-validation}

We validate attached gradients on two caustic-producing scenes (Figure~\ref{fig:gradient-validation}), again using a blurred-render target.
The Mirror scene (path length 4, 128~spp) differentiates the vertex positions of the conductor.
The Glass Panel scene (path length 4, 128~spp) differentiates the panel's normal map, exercising Jacobian transport at refraction.
All three variants again agree with the reference, which validates the Jacobian transport and the position adjoint.

\subsection{Memory Scaling}
\label{sec:results-memory}

We sweep maximum path length from 4 to 128 in the Guide scene (Figure~\ref{fig:depth-sweep}).
The guide is a straight glass pipe 320 widths long, and light reaches the screen after 12, 28, 60 and 124 reflections; by path lengths 16, 32 and 64, 25\%, 51\% and 71\% of the light arriving at 128 has reached it.
Naive AD grows linearly, reaching 11.38\,GB detached and 11.79\,GB attached at path length 16, and running out of memory at 32.
LRB-2-pass grows at the same rate from a lower constant: in the detached case, it reaches 10.53\,GB at 64 and fails at 128; when attached, it also stores a Jacobian and fails at 64.
Splitting its backward pass into chunks lowers the peak to 3.03\,GB detached and 4.47\,GB attached without changing the time (four chunks: 0.823\,s against 0.828\,s detached at path length 64, 0.829\,s against 0.822\,s attached at 32), but does not remove the per-vertex buffer.
\texttt{reslrb} and \texttt{lrb\_3pass} stay flat across the whole range, at 0.55\,GB detached, and attached 0.58\,GB for \texttt{lrb\_3pass} and 0.60\,GB for \texttt{reslrb}.

\begin{figure*}[t]
  \centering
  \includegraphics[width=\textwidth]{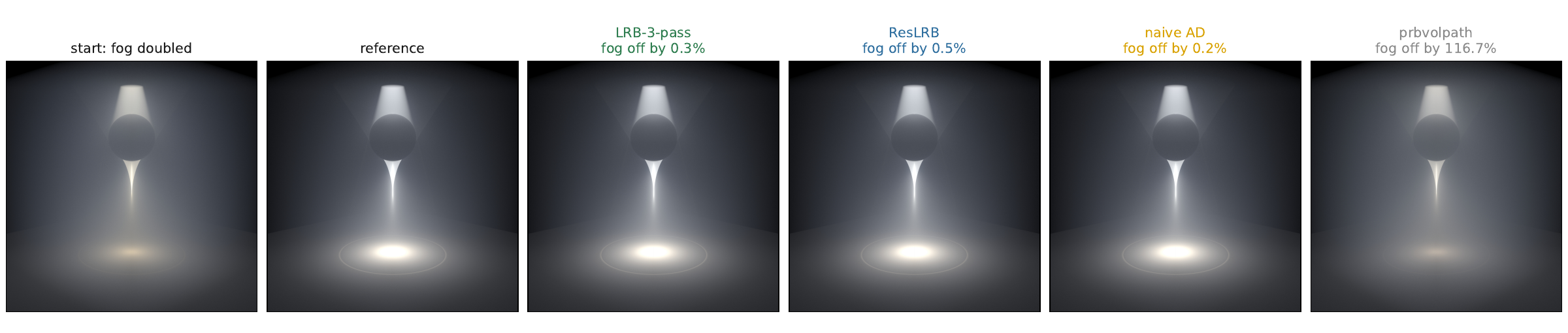}
  \caption{Recovering a fog's extinction from the caustic inside it, starting from twice the true value. The optimization runs at 160$^2$, 64~spp and path length 24; the panels are re-rendered at 2176$^2$ from the extinction each method recovered.}
  \label{fig:caustic-volume}
\end{figure*}

\begin{figure}[t]
  \centering
  \includegraphics[width=\columnwidth]{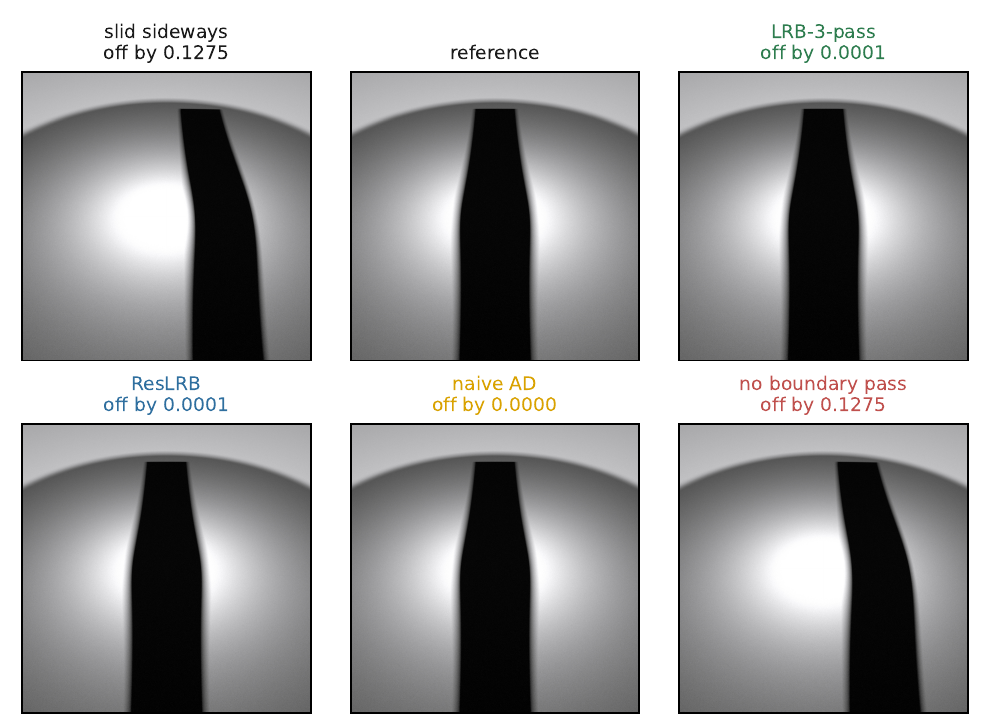}
  \caption{A glass sphere focuses light onto a floor, and a bar hanging out of frame casts a hard-edged stripe across the caustic. The bar is slid sideways and recovered from the image alone; its two edges carry the whole derivative. The optimization runs at $256^2$ and 256~spp with a path length of 6.}
  \label{fig:projective-opt}
\end{figure}

\begin{figure}[t]
  \centering
  \includegraphics[width=\columnwidth]{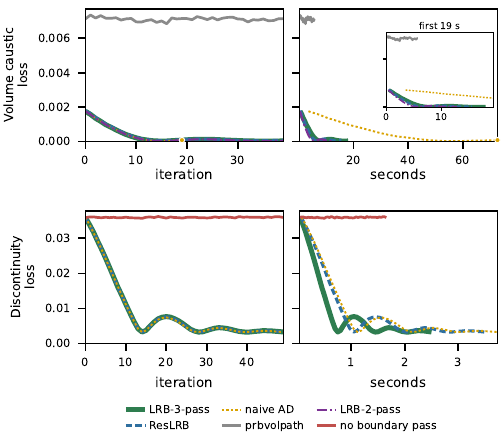}
  \caption{Loss per iteration (left) and per second (right) for both optimizations, at 256~spp for the discontinuity and 64~spp for the volume. Discontinuity row: first 50 iterations; the flat line has the boundary term off. Volume row: the inset shows the first seconds, and naive AD stops at 19 iterations, where its run hit our time limit.}
  \label{fig:loss-curves}
\end{figure}

\subsection{Timing}
\label{sec:results-timing}

Per-iteration cost is shown alongside memory in Figure~\ref{fig:depth-sweep} and across scenes in Figure~\ref{fig:rendering-stats-2x2}.
Naive AD is fastest only at path length 4 and is out of memory from 32 onwards.
Detached, \texttt{lrb\_3pass} rises from 0.016 to 0.164\,s between path lengths 4 and 128 and \texttt{reslrb} from 0.021 to 0.176\,s, both nearly flat beyond 64; \texttt{lrb\_2pass} reaches 0.323\,s.
Attached costs are higher and roughly track the number of vertices, since the ray Jacobian is built in both passes and chained and inverted at every vertex: \texttt{lrb\_3pass} rises from 0.047 to 1.683\,s and \texttt{reslrb} from 0.089 to 1.703\,s.
\texttt{lrb\_3pass} is the faster of the two at every path length here, detached and attached, with \texttt{reslrb} drawing level attached at 128.

\subsection{Participating Media}
\label{sec:results-volume}

Both methods extend to participating media; the volumetric formulation is given in the supplementary material.
Figure~\ref{fig:caustic-volume} shows them recovering a fog's extinction from the caustic inside it.
Both converge to within 0.5\% of the true extinction, while Mitsuba's \texttt{prbvolpath} does not reduce the loss.
Gradients agree with central finite differences to within 0.1\% for \texttt{lrb\_3pass} and 8.7\% for \texttt{reslrb}; at this sample count a single estimate of this gradient carries about 13\% relative noise for both, so neither figure measures a bias.
At $160^2$, 64~spp and path length 24, an iteration takes 0.47\,s for \texttt{lrb\_3pass\_vol} and 0.41\,s for \texttt{reslrb\_vol}, both using 0.57\,GB; naive AD takes 3.46\,s and 3.95\,GB.
LRB-2-pass is the fastest at 0.33\,s but uses 3.08\,GB.
Unlike on surfaces, \texttt{reslrb} is faster than \texttt{lrb\_3pass} here, since a medium scatters at every step and sensor connections are dense.
Figure~\ref{fig:loss-curves} shows the loss curves.
Naive AD follows the same descent at about seven times the cost per iteration; its run hit our time limit after 19 iterations, and we did not continue it.

\subsection{Discontinuities}
\label{sec:results-projective}

Figure~\ref{fig:projective-opt} repeats the recovery experiment on a visibility discontinuity, where the derivative comes entirely from the motion of the bar's two edges.
The optimization runs at $256^2$ and 256~spp with a path length of 6, and the boundary term draws 64 samples per pixel.
We shift the bar 0.128 units sideways, and LRB-3-pass, ResLRB and naive AD all move it back to within $10^{-4}$ units in about 40 iterations (Figure~\ref{fig:loss-curves}).
Without the boundary term the gradient is zero and the bar does not move.
The projective boundary term we use~\cite{Zhang2023Projective}, and how it is combined with the replay, are described in the supplementary material.

\section{Discussion}
\label{sec:discussion}

Table~\ref{tab:design-space} summarizes the four methods, which trade memory, gradient variance and time against one another.
No method wins everywhere; LRB-3-pass is the general frontrunner, and ResLRB may be the better choice when sensor connections are dense.

\textbf{When the constant-memory methods are needed.}
Naive AD stores the whole computation, so its memory grows with path length, sample count and resolution, and beyond the point where it runs out of GPU memory the constant-memory methods are the only option (Section~\ref{sec:results-memory}).
Below that point the choice is about speed.
On short paths naive AD is as fast or faster, since its graph is small and the replays retrace the path two or three times: the two are level on the detached Slab, and attached at path length 4 LRB-3-pass takes $1.7\times$ and $2.8\times$ the time of naive AD in the Mirror and Glass, because it carries the splat position through every replay.
As paths grow, writing and reading the graph dominates, and in the Guide at path length 16 detached LRB-3-pass takes $0.46\times$ the time of naive AD while attached it takes $1.3\times$.
LRB-2-pass, even chunked, still buffers every vertex, and we see little use for it outside debugging and validation.

\begin{table*}[t]
\centering
\small
\setlength{\tabcolsep}{4.5pt}
\definecolor{memgood}{RGB}{200,230,200}
\definecolor{membad}{RGB}{255,220,220}
\definecolor{vargood}{RGB}{200,230,200}
\definecolor{varbad}{RGB}{255,235,200}
\begin{tabular}{@{}l c c c c r r r r r r r@{}}
\toprule
 & & & & & \multicolumn{2}{c}{Attached} & \multicolumn{4}{c}{Detached} & Volume \\
\cmidrule(lr){6-7}\cmidrule(lr){8-11}\cmidrule(lr){12-12}
Method & Memory & Worst peak & Variance & Passes
 & Mirror & Glass & Slab & Room & Guide & Guide & Caustic \\
 & & & & & \tiny 4 & \tiny 4 & \tiny 4 & \tiny 16 & \tiny 16 & \tiny 128 & \tiny 24 \\
\midrule
\multicolumn{5}{@{}l}{\itshape What a light path does there} \\
\quad\itshape Connections per path & & & & & \itshape 0.03 & \itshape 0.27 & \itshape 0.09 & \itshape 0.61 & \itshape 0.24 & \itshape 0.73 & \itshape dense \\
\quad\itshape Surface vertices per path & & & & & \itshape 0.8 & \itshape 1.0 & \itshape 1.5 & \itshape 2.3 & \itshape 15.6 & \itshape 43.4 & \itshape --- \\
\midrule
\multicolumn{12}{@{}l}{Per-iteration time (s)} \\
Naive AD   & \cellcolor{membad}$O(k)$  & \cellcolor{membad}11.8\,GB & \cellcolor{vargood}$1\times$ & 1 & \textbf{0.078} & \textbf{0.117} & 0.027 & 0.135 & 0.154 & \emph{OOM} & 3.46 \\
LRB-2-pass & \cellcolor{membad}$O(k)$  & \cellcolor{membad}4.5\,GB  & \cellcolor{vargood}$1\times$ & 2 & 0.298 & 0.510 & 0.027 & 0.078 & 0.078 & 0.323 & \textbf{0.334} \\
ResLRB     & \cellcolor{memgood}$O(1)$ & \cellcolor{memgood}0.60\,GB & \cellcolor{vargood}$0.8$--$1.0\times$ & 2 & 0.371 & 0.550 & 0.034 & \textbf{0.064} & 0.079 & 0.176 & 0.406 \\
LRB-3-pass & \cellcolor{memgood}$O(1)$ & \cellcolor{memgood}0.58\,GB & \cellcolor{vargood}$1\times$ & 3 & 0.136 & 0.331 & \textbf{0.026} & 0.079 & \textbf{0.071} & \textbf{0.164} & 0.471 \\
\bottomrule
\end{tabular}
\vspace{0.6em}
\caption{The four methods. Left: memory scaling with path length $k$, worst measured peak, per-texel gradient variance relative to LRB-3-pass, and traversals. Right: per-iteration time (s) on a 16\,GB card at each scene's sample count from Section~\ref{sec:results} (32~spp Slab and Room, 128~spp Mirror and Glass, 64~spp Guide and Caustic), path length under each scene, fastest in bold. Italic rows: valid sensor connections and surface vertices per emitter path, from $10^5$ traced paths.}
\label{tab:design-space}
\end{table*}

\textbf{ResLRB or LRB-3-pass.}
LRB-3-pass has the same variance as naive AD, needs no tuning, and is the faster of the two constant-memory methods in every scene except the Room and the fog.
The choice depends on how many sensor connections a path makes.
Connections are rare in every scene: even the Guide at path length 128 averages only 0.73 per path and the Room 0.61, and in the Mirror 97\% of emitter paths make none, since they reach less than one surface vertex on average and those vertices are mostly specular.
The reservoir in ResLRB therefore almost always selects from zero candidates or one.
It pays off where connections are dense relative to path length, as in the Room (0.27 per vertex against 0.017 in the Guide at path length 128), where selecting one connection saves differentiating the rest and ResLRB is $1.2\times$ faster detached (0.064 against 0.079\,s).
Where connections are sparse, the third traversal of LRB-3-pass is cheap and LRB-3-pass is faster.

\textbf{How much variance ResLRB adds.}
Selecting one of several connections is a sampling decision, so its variance grows with the number of connections per path.
Where connections are sparse this is not measurable: per texel on the Guide normal map, ResLRB's gradient variance is $0.8$ to $1.0\times$ that of LRB-3-pass at every path length from 16 to 128, and across 29 scalar-parameter tests the median ratio is $0.96\times$.
Where connections are dense, as in the Room and the fog, the added noise is visible, and these are the scenes where ResLRB is faster.
In the fog it ends at a higher loss than LRB-3-pass but still recovers the extinction to within 0.5\%.
How much noise is acceptable depends on the application: where a fast approximate result is enough, as in exploratory or early optimization, ResLRB is the better choice, and where final accuracy matters, LRB-3-pass is.

\textbf{Beyond light tracing.}
The obstacle addressed here, recovering per-vertex adjoints during replay when one path reaches many image locations, arises in any integrator that breaks the one-path-one-pixel property.
Bidirectional methods, photon mapping and many-light techniques all produce such paths, and both stochastic compression and multi-pass accumulation apply to them.

\textbf{Limitations.}
Both constant-memory methods inherit the weaknesses of particle tracing: it is inefficient when most emitter paths miss the sensor, and it cannot find specular-diffuse-specular paths, which need methods that change the rendering algorithm itself~\cite{Xing2024DPMG}.
Like PRB, the replay handles only interior derivatives, and we add a boundary term for visibility discontinuities.
The projective term we use~\cite{Zhang2023Projective} was developed for camera-side path tracing.
Other boundary methods developed for path tracing~\cite{LiEdge2018, BangaruWarp2020, Xu2023WAS, Belhe2025Shaders, wu2025unbiased, Xu2026Kernel} may apply to light tracing in the same way, but this remains to be explored.
 
\section{Conclusion}
\label{sec:conclusion}

PRB's two-pass replay does not reach constant memory for light tracing: the per-vertex image-space adjoints that the backward replay needs cannot be reconstructed from constant-size per-sample state in one forward pass.
LRB-2-pass buffers the per-vertex splat data instead, which lowers the constant relative to naive AD but keeps memory linear in path length.

LRB-3-pass removes that scaling by adding a traversal that precomputes the adjoint-weighted radiance along each path.
It keeps every sensor connection and matches naive AD in expectation, and its attached formulation carries the splat-position adjoint through the replay in constant memory.
ResLRB reaches constant memory with two passes by keeping one reservoir-selected connection per path.
ResLRB is the faster of the two where sensor connections are dense, in participating media and in the detached Living Room.
Together with naive AD and LRB-2-pass, the two methods give four options that trade memory, variance and time, and the right choice depends on path length, connection density and the available memory.

\ifarxiv
\begin{acks}
We acknowledge with gratitude the funding provided for this research by Simon Fraser University as well as a Discovery Grant provided by the Natural Sciences and Engineering Research Council of Canada. We are also grateful to the anonymous reviewers of an earlier version of this paper, who suggested the LRB-3-pass formulation. Additionally, we thank Benedikt Bitterli \cite{resources16} for providing the Living Room scene.
\end{acks}
\else
\section*{Acknowledgments}
We are grateful to the anonymous reviewers of an earlier version of this paper, who suggested the LRB-3-pass formulation. Funding and scene credits are withheld for review.
\fi

\bibliographystyle{ACM-Reference-Format}
\bibliography{references}

\end{document}